\documentclass[]{aa}
\usepackage[varg]{txfonts}
\usepackage{graphicx}
\usepackage{textcomp}
\usepackage{color}
\usepackage{bbold}
\usepackage{breqn}
\usepackage{amsmath}
\usepackage{pstricks} 
\usepackage{bm} 
\usepackage{relsize}
\usepackage[colorlinks=true,urlcolor=blue,citecolor=blue,linkcolor=blue,breaklinks=true]{hyperref}
\usepackage[]{hyperref}
\usepackage{twoopt}

\def\xlinkspace#1 #2{%
 \ifx\relax#2%
 \xlinkdash#1-\relax
 \else
 \xlinkdash#1 -\relax
 \expandafter\xlinkspace\expandafter#2%
 \fi}

\def\xlinkdash#1-#2{%
 \ifx\relax#2%
 \tmp{#1}%
 \else
 \tmp{#1-}%
 \expandafter\xlinkdash\expandafter#2%
 \fi}

\bibpunct{(}{)}{;}{a}{}{,} 

\makeatletter

 \newcommandtwoopt{\citeads}[3][][]{%
   \nonstopmode
   \href{https://ui.adsabs.harvard.edu/abs/#3}%
        {\def\hyper@linkstart##1##2{}%
         \let\hyper@linkend\@empty\citealp[#1][#2]{#3}}
   \biblink{#3}{\href{https://ui.adsabs.harvard.edu/abs/#3}{ADS}}%
   \errorstopmode}            
   
 \newcommandtwoopt{\citepads}[3][][]{%
   \nonstopmode
   \href{https://ui.adsabs.harvard.edu/abs/#3}%
        {\def\hyper@linkstart##1##2{}%
         \let\hyper@linkend\@empty\citep[#1][#2]{#3}}
   \biblink{#3}{\href{https://ui.adsabs.harvard.edu/abs/#3}{ADS}}
   \errorstopmode}            
   
 \newcommandtwoopt{\citetads}[3][][]{%
   \nonstopmode
   \href{https://ui.adsabs.harvard.edu/abs/#3}
        {\def\hyper@linkstart##1##2{}%
         \let\hyper@linkend\@empty\citet[#1][#2]{#3}}
   \biblink{#3}{\href{https://ui.adsabs.harvard.edu/abs/#3}{ADS}}%
   \errorstopmode}            
   
 \newcommandtwoopt{\citeyearads}[3][][]{%
   \nonstopmode
   \href{https://ui.adsabs.harvard.edu/abs/#3}%
        {\def\hyper@linkstart##1##2{}%
         \let\hyper@linkend\@empty\citeyear[#1][#2]{#3}}
   \biblink{#3}{\href{https://ui.adsabs.harvard.edu/abs/#3}{ADS}}%
   \errorstopmode}            
\makeatother

\makeatletter
\newcommand{\bibnote}[2]{\@namedef{#1note}{#2}}
\newcommand{\biblink}[2]{\@namedef{#1link}{#2}}
\makeatother

\newcommand{\be}{\begin{equation}}
\newcommand{\ee}{\end{equation}}

\newcommand{\bea}{\begin{eqnarray}}
\newcommand{\eea}{\end{eqnarray}}

\begin{document}

\title{A method for global inversion of multi-resolution solar data}
\author{
  J. de la Cruz Rodr\'{i}guez
}

\offprints{J. de la Cruz Rodr\'iguez \email{jaime@astro.su.se}}

\institute{
Institute for Solar Physics, Dept. of Astronomy, Stockholm University, AlbaNova University Centre, SE-106 91 Stockholm, Sweden
}
\titlerunning{A global inversion method}
\authorrunning{de la Cruz Rodr\'iguez}

\date{Received; Accepted }

\abstract 
{
Understanding the complex dynamics and structure of the upper solar atmosphere benefits strongly from the use of a combination of several diagnostics. Frequently, such diverse diagnostics can only be obtained from telescopes and/or instrumentation operating at widely different spatial resolution. To optimize the utilization of such data, we propose a new method for the global inversion of data acquired at different spatial resolution. The method has its roots in the Levenberg-Marquardt algorithm but involves the use of linear operators to transform and degrade the synthetic spectra of a highly resolved guess model to account for the the effects of spatial resolution, data sampling, alignment and image rotation of each of the data sets. 

We have carried out a list of numerical experiments to show that our method allows extracting spatial information from two simulated datasets that have gone through two different telescope apertures and that are sampled in different spatial grids. Our results show that each dataset contributes in the inversion by constraining information at the spatial scales that are present in each of the datasets, without any negative effects derived from the combination of multiple resolution data.

This method is especially relevant for chromospheric studies that attempt at combining datasets acquired with different telescopes and/or datasets acquired at different wavelengths, both limiting factors in the resolution of solar instrumentation. The techniques described in the present study will also help addressing the ever increasing resolution gap between space-borne missions and forthcoming ground-based facilities.
}

    \maketitle

\section{Introduction} \label{sec:intro}
\begin{figure*}
\centering
\includegraphics[width=\textwidth]{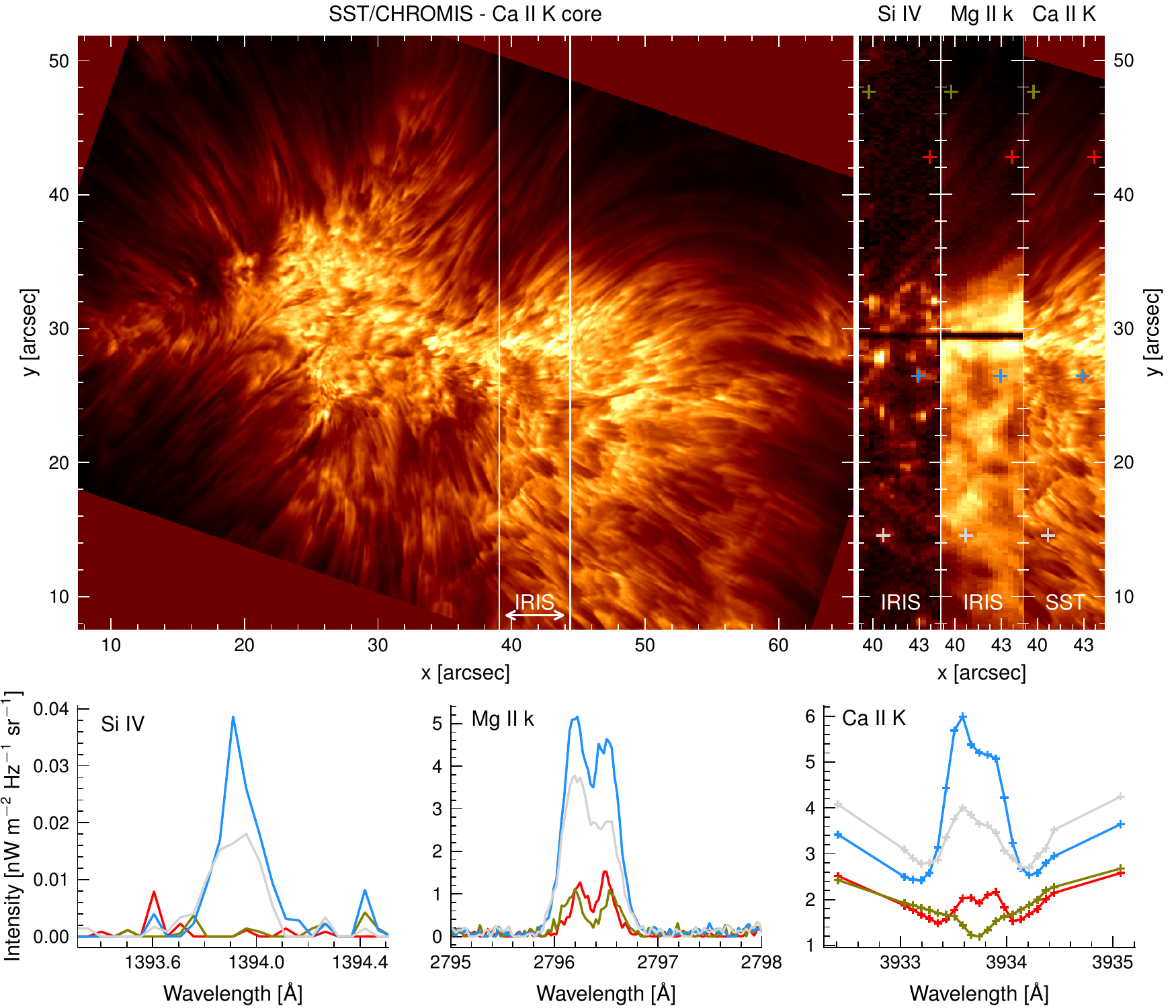}
\caption{Chromospheric plage observation acquired with the CHROMIS instrument at the Swedish 1-m Solar Telescope and with NASA's IRIS satellite on 2016-09-14 at 09:06 UT. The upper-left panel illustrates the CHROMIS field-of-view where the IRIS raster scan has been indicated. The three upper rightmost panels depict the overlapping FOV for a given time-step in the core of the \ion{Si}{iv}~1400~\AA\ line, the \ion{Mg}{ii}~k line and the \ion{Ca}{ii}~K line. The colored crosses indicate the location of example spectra, shown in the bottom row with the same color coding for each line.}
\label{fig:intro}
\end{figure*}
Inversion techniques allow inferring solar physical parameters from the reconstruction of spectropolarimetric observations by assuming a model atmosphere: Milne-Eddington, constant slab, depth-stratified. One underlying assumption in inversion methods is that the model includes all physical and instrumental ingredients that are required in order to model the observational data. In the vast majority of cases, inversions have been performed in a 1D fashion, where each pixel is assumed to be independent from the rest and processed individually. This assumption makes it very easy to define a model atmosphere that can reproduce the observed spectral lines reasonably well, but the instrumental implications are far more severe.

The fidelity of inversions greatly depends on the diagnostic potential of the spectral lines under consideration. When the considered selection of spectral lines are sensitive to a range of heights from the solar atmosphere, depth-stratified inversions have allowed for the reconstruction of vertical gradients of these parameters in certain parts of the solar atmosphere. The inclusion of more spectral lines in the inversions usually translates into better depth-resolution and better constrained output models (e.g., \citeads{2018A&A...620A.124D}; \citeads{2019A&A...622A..36R}; \citeads{2019A&A...627A.101V}).

While observing many photospheric lines simultaneously can be achieved by acquiring data towards the ultraviolet (UV), there are very few spectral lines that sample the chromosphere and they are spread over a very large wavelength range, including lines in the UV (L$\alpha$, \ion{Mg}{ii}~h\&k), visible (\ion{Ca}{ii}~H\&K, \ion{He}{i}~D$_3$, H$\alpha$) and infrared (\ion{Ca}{ii}~IR triplet lines, \ion{He}{i}~10830~\AA). Therefore, in most cases the instrumental requirements to observe each of these lines leads to data of different nature in spectral coverage, cadence and spatial resolution. The simultaneous interpretation of such combination of datasets is far from trivial, leaving aside the challenge of including all relevant physical descriptions that might be needed to properly model these diagnostics: non-local thermodynamical equilibrium (NLTE), time-dependent ionization, 3D radiative transfer.

Fig.~\ref{fig:intro} shows a simultaneous chromospheric observation of a plage target acquired in the \ion{Ca}{ii}~K line with the CHROMIS instrument at the Swedish 1-m Solar Telescope (SST, \citeads{2003SPIE.4853..341S}) and in the \ion{Mg}{ii}~k and \ion{Si}{i}~140~nm lines acquired with NASA's IRIS satellite (\citeads{2014SoPh..289.2733D}). While the CHROMIS data are acquired at a spatial resolution $dx=dy=0.0375\arcsec$ and a time-cadence of $\Delta t=6$~s, the IRIS diagnostics can sample higher layers of the chromosphere and the transition region, but at lower spatial resolution $(dx,dy)=(0.33\arcsec,0.165\arcsec)$ and a cadence of $\Delta t=19$~s. There have been attempts to combine similar ground-based and space-borne datasets in a simultaneous inversion (\citeads{2019A&A...627A.101V}) by resampling them to the same spatial grid, but due to the difference in spatial resolution of each spectral window, it has been challenging to find a good match for all diagnostics as instrumental effects were not properly addressed.

Recent developments to the Levenberg-Marquardt algorithm, proposed by \citetads{2012A&A...548A...5V}, have allowed correcting for the contrast-decreasing effect of the telescope point-spread-function (PSF) during the inversion. As a consequence of this inclusion, the inversion problem became a global problem, including simultaneously all pixels in the field-of-view (FOV), which are spatially-coupled by the telescope PSF. A list of excellent publications that targeted the physics of the solar photosphere made use of the latter technique to study in detail plage (\citeads{2015A&A...576A..27B}; \citeads{2019arXiv190807464B}), sunspots (\citeads{2013A&A...557A..24V}; \citeads{2013A&A...557A..25T}; \citeads{2014A&A...568A..60L}) and quiet-Sun (\citeads{2016A&A...593A..93D}; \citeads{2016A&A...594A.103D}) using Hinode SOT observations. One limitation in the implementation of \citetads{2012A&A...548A...5V} is that both the data and the model must be given in the same discrete spatial grid and, at least in their applications, they only consider one degradation mechanism. A simplified alternative technique based on the sparsity of the data and spatial regularization was also proposed by \citetads{2015A&A...577A.140A}.

Global inversion methods make use of information from neighbouring pixels to constrain the solution in one location, which makes the problem less sensitive to parameter degeneracy and to the presence of noise. In traditional 1D pixel-to-pixel inversions, that constraint is ignored and the problem is solved freely in relation to other pixels.

A different approach has been to account for pixel straylight effects (see \citeads{2019A&A...626A..55S}) as a pre-processing step, combined with traditional pixel-to-pixel 1D inversions. The datasets were deconvolved using a PSF that described the degradation process before performing the inversion. Such techniques allowed \citetads{2011Sci...333..316S}, \citetads{2011ApJ...734L..18J}, \citetads{2013A&A...549L...4R}, \citetads{2013A&A...553A..63S}, \citetads{2016MNRAS.460..956Q}, \citetads{2016MNRAS.460.1476Q}, \citetads{2017A&A...601L...8B} and \citetads{2017ApJ...849....7O} (among others) to achieve similar fidelity in their photospheric inversions to a spatially-coupled method, because they could recover similar levels of contrast in their reconstructed models. However, the latter do not treat the inversion as a global problem, and they are potentially more affected by parameter degeneracy and noise.

The application of similar techniques has not been attempted with chromospheric data. In order to properly constrain chromospheric inversions, the inclusion of more than one spectral line and/or atomic species are desired (\citeads{2018A&A...612A..28L}; \citeads{2019ApJ...870...88E}; \citeads{2019A&A...627A.101V}). But in order to do so, it is almost inevitable combining data acquired at very different wavelengths or combining ground-based observations with satellite observations and therefore instrumental issues and spatial resolution discrepancies will be present. All the aforementioned global inversion methods do not allow, in their present form, dealing with data of different resolution.

We propose a generalization of the Levenberg-Marquardt algorithm, based on the ideas of \citetads{2012A&A...548A...5V}, for the inversion of data acquired at different spatial resolutions, while including different degradation and calibration processes in the form of linear operators. In this approach, each dataset contributes to the inversion by constraining the spatial scales that are present in the data in a self-consistent way.

Additionally, we have included spatial Tikhonov regularization in the inversion algorithm to improve the convergence speed (number of iterations) and the fidelity of the inversions. Our formalism decouples the spatial grid of the model from that of the observations, allowing to provide the model in a denser spatial grid than that of the observations, which can be used to diminish the effect of pixel discretization and high-frequency damping close to the Nyquist cut-off limit.

\section{Spatial constraints for solar inversions}\label{sec:constraints}
The first spatially-coupled inversions were introduced by \citetads{2012A&A...548A...5V}, who utilized the PSF of the telescope to couple the synthetic spectra from different pixels before comparing them with the observations. The latter approach transforms the inversion into a global minimization problem with as many parameters as $n_{\mathrm{pix}} \times n_{\mathrm{par}}$. 

However, other forms of spatial constraints can be applied to inversions. In many stellar applications, spatial parameter degeneracies and convergence issues have been addressed by using regularization techniques like Tikhonov regularization or maximum entropy regularization (e.g., \citeads{1990A&A...230..363P}; \citeads{2002A&A...381..736P}). The idea behind regularization is to impose certain mathematical constraints that help converging the inversion and to discard physically unlikely solutions, based for example on smoothness. While the origin of regularization is different than applying a PSF to couple the solutions of different pixels, including the former also transforms the inversion problem into a global problem, so in principle both techniques can be used simultaneously.

The Levenberg-Marquardt algorithm (LMA) has been extensively used in solar applications, because it shows excellent convergence properties compared to other gradient-descent methods. The algorithm can be very easily modified to include $l-2$ regularization (see derivation in e.g., \citeads{2019A&A...623A..74D}). The latter is introduced in the form of a squared penalty function that contributes to the definition of $\chi^2$ and which has the effect of discouraging certain solutions in the fitting process. However, \citetads{2019A&A...623A..74D} only applied regularization within the parameters of a pixel (as a function of optical-depth) and not to set spatial constraints in the parameters of the model.

In the following we propose how to use both techniques, regularization and spatial instrumental degradation, in a general global problem. Using a similar notation as \citetads{2019A&A...623A..74D}, the merit function $\chi^2$ is modified to include a sum of penalty terms $\Gamma_k(\bm{p})$ that depend on the current estimate of the model parameters $\bm{p}$:
\begin{equation}
  \chi^2 = \sum_{d=1}^{N_d} \left(\frac{o_d - s_d(\bm{p})}{\sigma_d}\right)^2 + \sum_{k=1}^{N_p} \alpha_k \Gamma_k(\bm{p})^2,\label{eq:chi2}
\end{equation}
where $o_d$ is the d-$th$ observed data-point, $s_d$ is the corresponding prediction of our model, $\sigma_d$ is the estimate of the noise, $\alpha_k$ is the weight of the regularization function $\Gamma_k(\bm{p})$.

The first term in Eq.~(\ref{eq:chi2}) depends on the data and the predictions of our model, whereas the second term only depends on the model parameters, but not on the actual data. Therefore, a proper normalization of the model parameters is required in order for this method to work. The modified Levenberg-Marquardt algorithm computes the model corrections by solving the following system of equations:
\begin{equation}
  \bigg(\mathbf{A} + \lambda\mathbf{I}\mathbf{A}\bigg)\bm{\Delta p} = \mathbf{J}^T\bm{r} - \mathbf{L}^T \bm{\mathit{\Gamma}} \equiv \bm{b}, \label{eq:lmr}
\end{equation}
where $\mathbf{A}$ is the linearized total Hessian matrix, $\mathbf{I}$ is the identity matrix, $\bm{\Delta p}$ is the correction to the model parameters, $\mathbf{J}$ is the Jacobian of the model prediction and $\mathbf{L}$ is the Jacobian of the penalty functions and the $\lambda$ parameter can be used to stabilize the predicted solution of the system.   $\bm{\mathit{\Gamma}}$ and $\bm{r}$ are penalty functions and the residue in vector form. The total Hessian matrix has two contributions:
\begin{equation}
  \mathbf{A} = \mathbf{J}^T\mathbf{J} + \mathbf{L}^T\mathbf{L}.
\end{equation}

So far there is no indication in Eq.~(\ref{eq:chi2}) and (\ref{eq:lmr}) of any explicit spatial coupling and therefore they can be used generally in any situation. Obviously the structure of those matrices and vectors change according to the nature of the problem.

\subsection{The global inverse problem}\label{sec:glob}
\begin{figure}
\centering
\includegraphics[width=0.85\columnwidth]{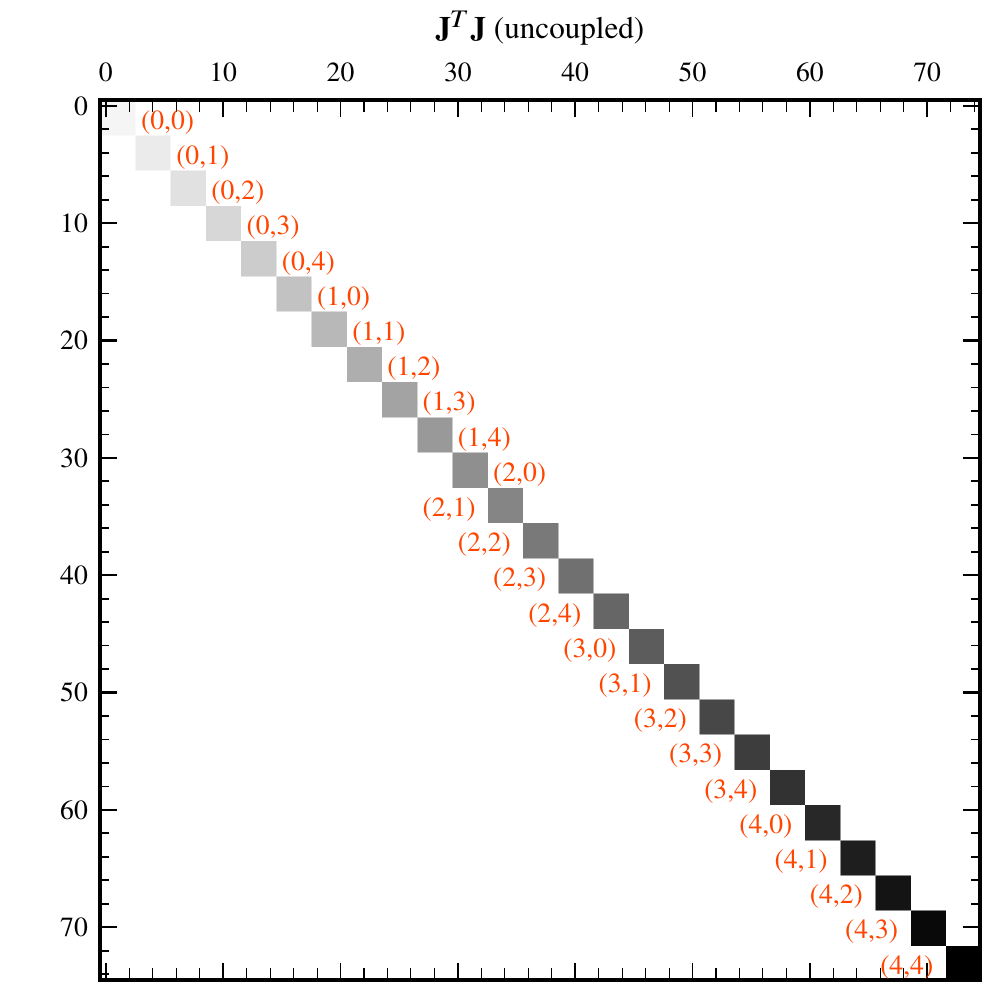}
\caption{Structure of the global Hessian matrix without spatial contraints for a case with $n_x= n_y=5$ and $n_{\mathrm{par}}=3$. The color coding indicates the subspace containing parameters from the same pixel. The coordinates  $(y,x)$ of the pixel associated with each subspace are indicated next to  each subspace. Each block has the dimension of the number of parameter per pixel, in this example $3\times 3$.}
\label{fig:blkdiag}
\end{figure}

Without including spatial contraints of any sort, the Hessian matrix $\mathbf{J}^T\mathbf{J}$ would look like a block-diagonal matrix where each block contains the Hessian of one pixel (see Fig.~\ref{fig:blkdiag}). Therefore we can treat each block independently, performing a traditional pixel-to-pixel inversion by splitting the resolution of this linear system in smaller problems. The inclusion of spatial constraints adds couplings between these blocks/subspaces in the form of bands in the Hessian matrix as shown by \citet{2012A&A...548A...5V}. We will see that imposing Tikhonov regularization leads to the addition of very localized bands to the Hessian matrix, whereas including the effect of a PSF leads to more extended bands.  In the following sections we will preserve the notation and structure of the matrix shown in Figure~\ref{fig:blkdiag}. 

\subsubsection{Tikhonov regularization}\label{sec:tik}
The idea behind Tikhonov regularization \citep{Tikhonov77} is to impose certain level of smoothness in the solution of the LM algorithm. In principle we can attain that effect by penalizing very steep gradients from the solutions of neighboring pixels, but other forms of regularization function can be used in this formalism. For each pixel we can define two penalty functions per type of physical parameter, such as:
\begin{eqnarray}
  \Gamma_{(y,x)}^0 &=& (p_{y,x} - p_{y-1,x}), \\
  \Gamma_{(y,x)}^1 &=&(p_{y,x} - p_{y,x-1}). 
\end{eqnarray}

\begin{figure}
\centering
\includegraphics[width=0.85\columnwidth]{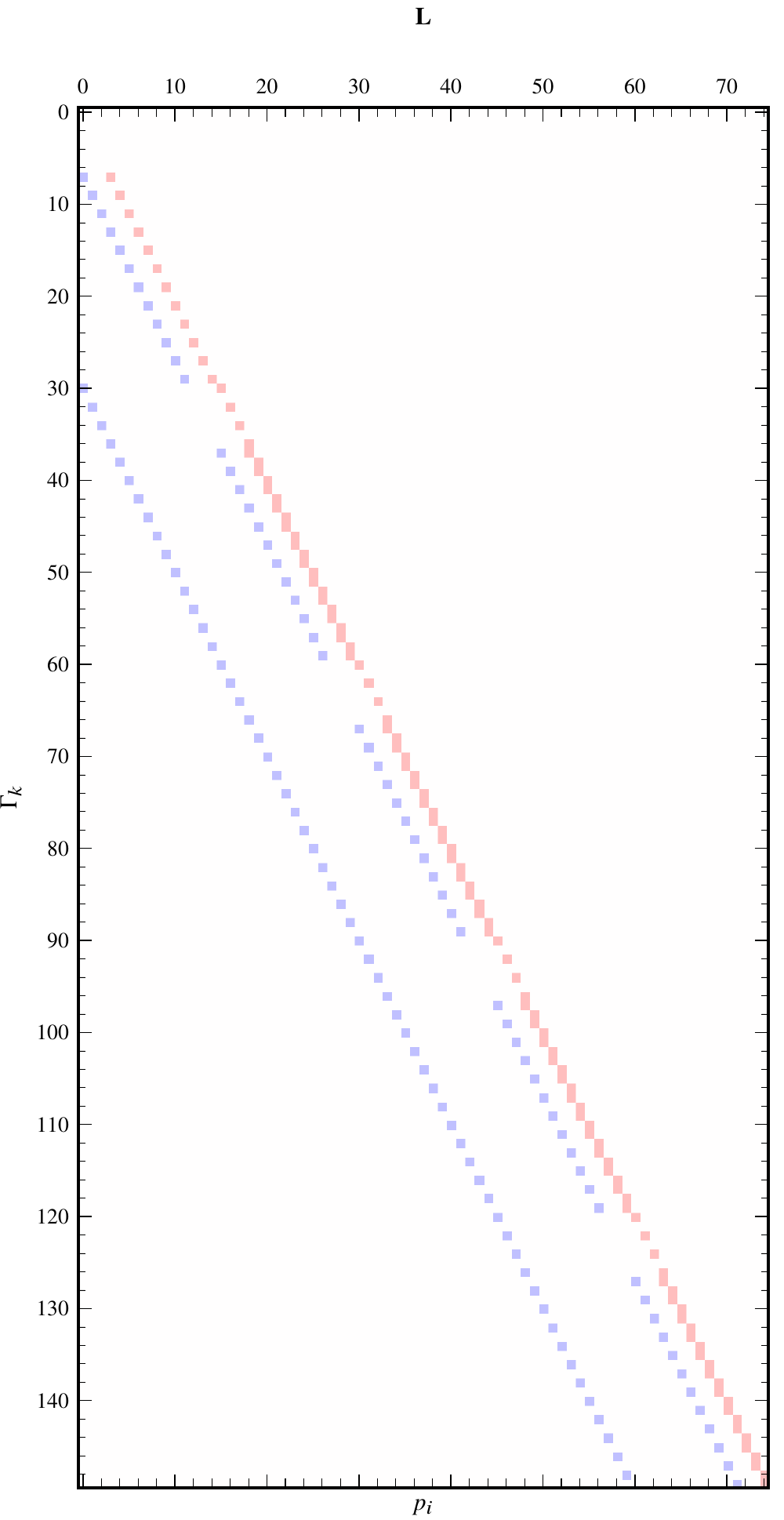}
\includegraphics[width=0.85\columnwidth]{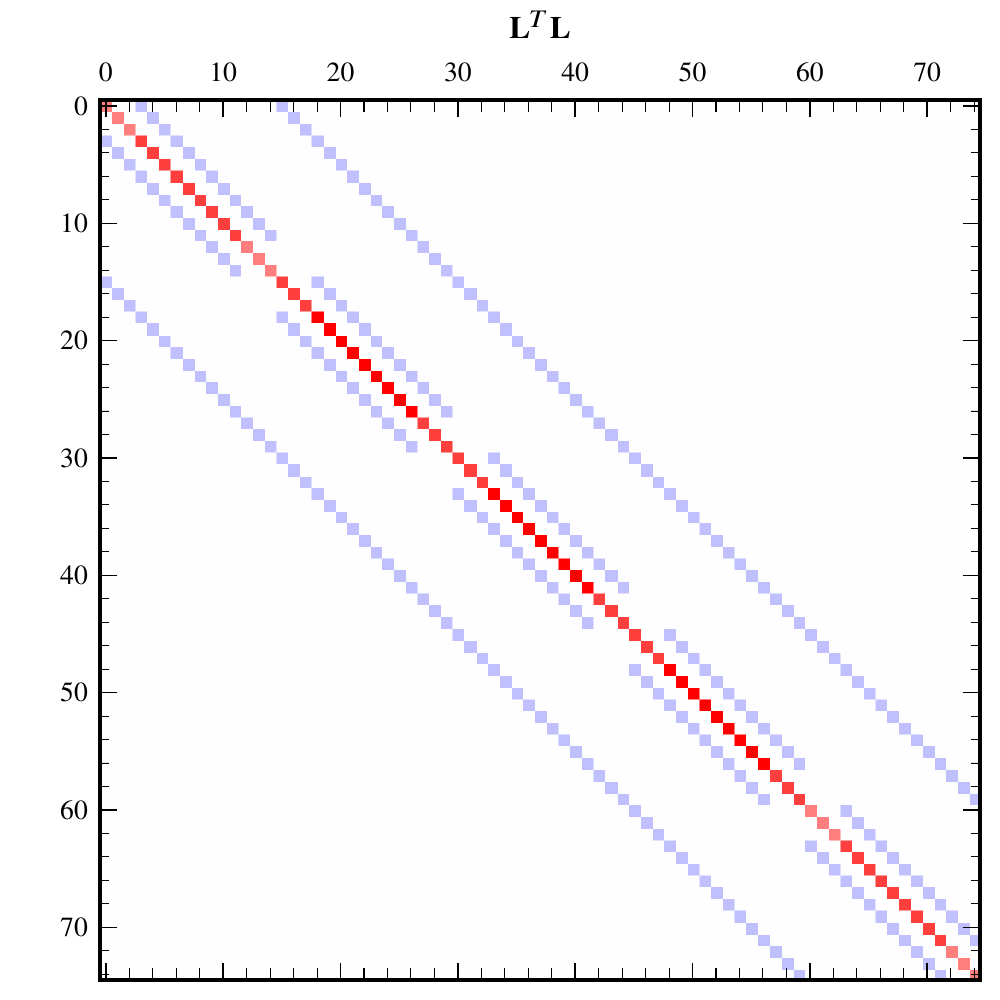}

\caption{\emph{Top:} The Jacobian of the regularization functions. The regularization functions are ordered along the $y$-axis. The free parameters for all pixels are displayed along the $x$-axis. \emph{Bottom:} The contribution to the Hessian from the spatial regularization functions.}\label{fig:tik}
\label{fig:LL}
\end{figure}

In that case, the corresponding $\mathbf{L}^T\mathbf{L}$ matrix would have in each row a contribution of 4 in the diagonal that originates from the cross product of $p_{y,x}$, two contributions shifted by $\pm n_{\mathrm{par}}$ elements of -1 and another two -1 contributions shifted by $\pm n_{\mathrm{par}}\times n_x$ elements originating from the cross product of $p_{y-1,x}$. The structure of this matrix is shown in Fig.~\ref{fig:tik}. The gaps in the matrix originate from the edges of the images, where some of the regularization functions cannot be evaluated. This effect is perhaps more obvious in the representation of $\mathbf{L}$. For example in the upper part of that matrix, there are many pixels where the regularization function cannot be defined for coordinates $(x,y-1)$ and both regularization functions are undefined for $(x,y)=(0,0)$. 

By adding $\mathbf{L}^T\mathbf{L}$ to the unconstrained Hessian matrix, we have effectively coupled the parameters from different pixels. Once the $\mathbf{L}$ matrix is known, the right-hand side regularized contribution is trivially computed with a matrix-vector multiplication with the $\bm{\mathit{\Gamma}}$ vector. The system of equations is largely sparse though and obtaining the solution is relatively fast and trivial with any iterative method like Conjugate Gradient (CG) or Biconjugate gradient stabilized method (BiCGStab).

\subsubsection{Instrumental effects}\label{sec:psf}

Including the effects of a spatial PSF or any other linear operator is not as trivially done (\citeads{2012A&A...548A...5V}). Usually, the effect of the PSF is to add many off-diagonal blocks that make the system less sparse, slower to construct and slower to solve. Although the method was explained in great detail by van Noort, for completeness we summarize some aspects of our implementation.

\begin{figure}
\centering
\includegraphics[width=0.85\columnwidth]{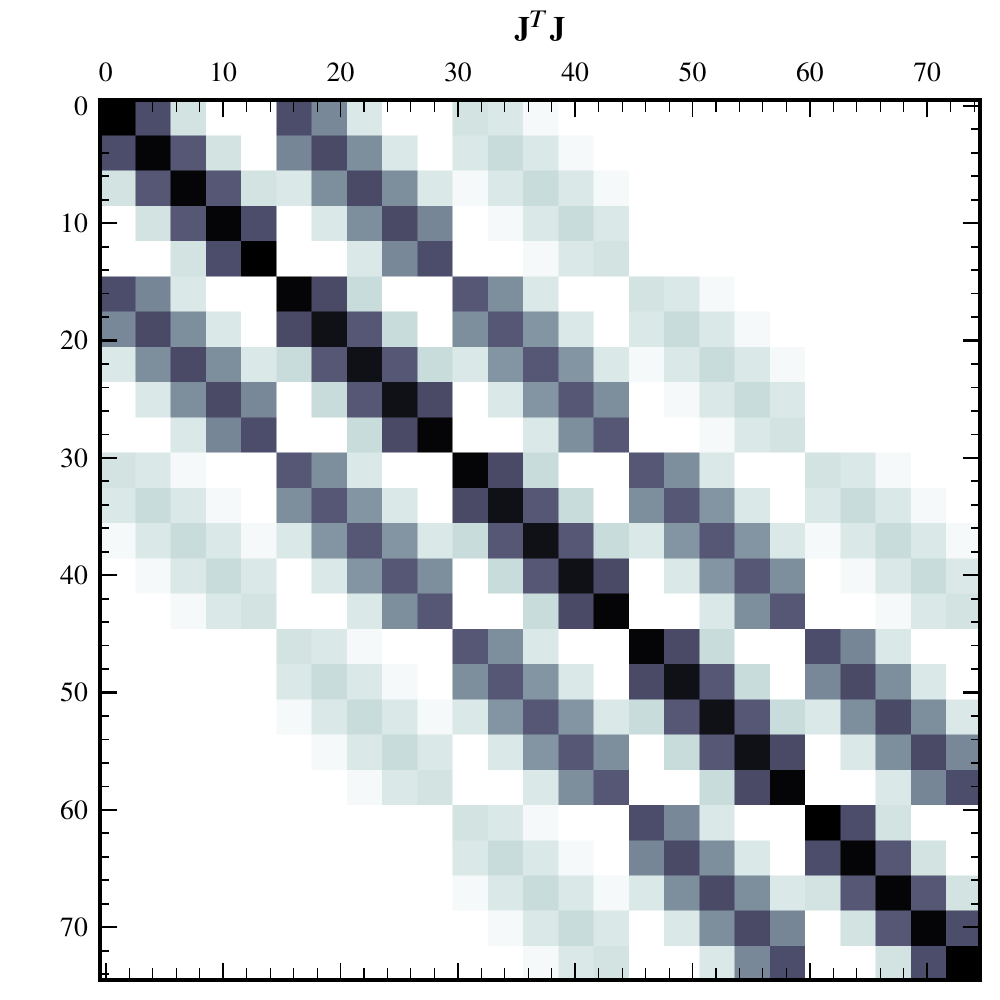}
\caption{The spatially coupled Hessian resulting from applying a $3\times 3$ pixels PSF of $\mathrm{FWHM}=2$~pixels. Each coloured square corresponds to a subspace of $n_{\mathrm{par}} \times n_{\mathrm{par}}$ where we have set all elements to 1 for illustration purposes. In reality the latter would also be scaled according to the products of response functions within each subspace. The relative scaling therefore is only due to the $Y$ term that multiplies each subspace.} \label{fig:PSF}
\end{figure}

Following the \citetads{2012A&A...548A...5V} approach, the structure of the Hessian matrix would contain cross-products of individual Jacobians from different pixels, weighted with the autocorrelation function of the PSF ($Y=\varphi * \varphi$), which appears from the matrix product of the transposed spatially coupled Jacobian times itself non-transposed\footnote{The autocorrelation function can be represented as the product of the convolution operator by its transposed form, which is contained in the spatially coupled Jacobians product.}:
\begin{multline}
  \mathbf{\bar{J}}^T\mathbf{\bar{J}} = 
  \begin{bmatrix}
    Y_{0,0} J_{0,0}^T J_{0,0} & Y_{0,1} J_{0,0}^T J_{0,1} & \dots & Y_{n-1,m-1} J_{0,0}^T J_{n-1,m-1} \\
    Y_{0,-1} J_{0,1}^T J_{0,0} & Y_{0,0} J_{1,1}^T J_{1,1} & \dots & Y_{n-2,m-2} J_{0,0}^T J_{n-1,m-1} \\
    \vdots & \vdots & \ddots & \vdots \\
     \ & \dots & &  Y_{0,0} J_{n-1,m-1}^T J_{n-1,n-1}
  \end{bmatrix},
  \label{eq:matpsf}
\end{multline}
where $\mathbf{\bar{J}}$ is the spatially coupled Jacobian and the subindexes $(m,n)$ both span over the $n_y\cdot n_x$ elements of the map.
In Eq.~(\ref{eq:matpsf}), the subindexes of the autocorrelation function $Y$ are relative to its center.
Figure~\ref{fig:PSF} shows an example of the structure of such matrix using a relatively small Gaussian PSF of $3\times 3$ pixels. The reader is referred to \citetads{2012A&A...548A...5V} for more examples with larger FOVs.

By comparing Fig.~(\ref{fig:LL}) and (\ref{fig:PSF}), it becomes quite obvious that dealing with the effect of a PSF involves a more complex linear system that is less sparse than the regularization contribution. Solving this system of equations by direct methods is very slow or even unrealistic for very extended PSFs because the inverse of the Hessian is not sparse. Instead in these applications we have used the BiCGStab solver implemented in the Eigen-3 library for sparse linear algebra \citep{eigenweb}, which solves the system iteratively without ever computing explicitly the inverse of the Hessian matrix.

\subsection{Generalization of the problem}\label{sec:gen}
One aspect that \citetads{2012A&A...548A...5V} did not explore, was the utilization of different PSFs for different spectral ranges. That would be the case of a simultaneous observation of a solar target with different facilities or instruments at different spatial resolutions. In order to do so, we can simply redefine our merit function as a sum of individual merit functions, one per spatial PSF, as follows:
\begin{dmath}
\bar{\chi}^2(\bm{p}) = \frac{1}{N_n}\sum_{d=1}^{N_1 }\left(\frac{o_d - P_1* s_d(\bm{p})}{\sigma_d}\right)^2 + \frac{1}{N_n}\sum_{d=N_1+1}^{N_2 }\left(\frac{o_d - P_2* s_d(\bm{p})}{\sigma_d}\right)^2{+ \dots +} + \frac{1}{N_n}\sum_{d=N_{n-1}+1}^{N_n}\left(\frac{o_d - P_n* s_d(\bm{p})}{\sigma_d}\right)^2+  \sum_{k=1}^{N_p} \alpha_k \Gamma_k(\bm{p})^2=\frac{1}{N_n}\sum_{r=1}^{n}\left[\sum_{d=N_{r-1}+1}^{N_r }\left ( \frac{o_d - P_r* s_d(\bm{p})}{\sigma_d}\right)^2 \right]+  \sum_{k=1}^{N_p} \alpha_k \Gamma_k(\bm{p})^2, \label{eq:chi2new}
\end{dmath}
where the $r$ subindex extends over all observed windows with a unique spatial PSFs and the subindex $d$ extends over the data points that correspond to each region. $P_r$ is the corresponding PSF of region $r$, that operates over the synthetic profiles.
The resulting LM corrections are computed via Eq.~(\ref{eq:lmr}), with the exception that now the $\mathbf{A}$ matrix is a sum of sparse Hessians and the RHS ($\bm{b}$) is computed as a sum of residues multiplied by the corresponding transposed spatially-coupled Jacobian ($\bar{\mathbf{J}}_{r}^{T}$):
\begin{eqnarray}
	\mathbf{A} &=& \mathbf{L}^T\mathbf{L} +  \sum_{r=1}^{n} \bar{\mathbf{J}}_{r}^T \bar{\mathbf{J}}_{r}, \\
	\bm{b} &=& - \mathbf{L}^T\bm{\mathit{\Gamma}}+\sum_{r=1}^{n} \bar{\mathbf{J}}_{r}^T \bm{r}_r .\label{eq:residue}
\end{eqnarray}

However, we do not need to restrict ourselves to applying a point-spread-function. Any degradation operation that can be written as a linear operator can be applied to the synthetic data. 
If we consider the synthetic map at each wavelength, packed in a vector ($\bm{s}$), we can write the degradation operation ($\mathbf{D}$) as a matrix-vector product, or more generally, a chain of operators describing each of the degradation steps applied to the data: telescope PSF, $(x,y)$-shift, rebinning, bilinear interpolations, etc.
\begin{equation}
	\bm{s}_{deg} =  \mathbf{D} \cdot \bm{s}= \left [\mathbf{D}_1 \mathbf{D}_2 \dots \mathbf{D}_n \right ]\cdot \bm{s}.
\end{equation}

Another insight from Eq.~(\ref{eq:chi2new}) is that in principle the parameters of our model ($\bm{p}$) and the observables do not need to be given in the same grid. In fact, each of the spectral windows with a unique spatial degradation operator could be given in a different spatial grid. We note  however that $\mathbf{\bar{J}}$ is given in the grid of the model, but $\bm{r}$ should be given in the degraded grid. Therefore a priori it is not obvious how to compute the RHS of Eq.~(\ref{eq:lmr}) where we have a product of $\mathbf{\bar{J}}^T$  and $\bm{r}$.  We will see that working with linear operators naturally addresses the conversion from different spatial grids. Obviously, the model grid must be sufficiently fine to allow performing the inversion of the highest resolution data that are included.
In \S\ref{sec:oper} we will discuss the structure of some of these operators.

 \subsection{Degradation operators}\label{sec:oper}
Working with explicit linear operators in sparse form requires some extra book keeping, but there are some advantages too:
\begin{enumerate}
\item the computation of the autocorrelation function in all pixels is trivially achieved by computing $\mathbf{D}^T\mathbf{D}$.
\item when we include rebinning/resampling operators, the data can be given in a different spatial grid than the model. The transposition of the degradation operator naturally gives how to relate $\bar{\mathbf{J}}^T$, which is given in the grid of the model atmosphere, to the spatial grid of the residue ($\bm{r}$) in the right-hand term (Eq.~\ref{eq:residue}).
\end{enumerate}

In the following, we will illustrate degradation operator matrices that operate over one individual 2D maps. The same operator must be applied to all corresponding points in the spectral direction for the data and to the terms of the spatially-coupled Jacobian. Each block in the Hessian matrix is affected by the same value of the autocorrelation of the operator.


\subsubsection{Convolution with a point-spread-function}\label{sec:psf}
The convolution of the data with a spatial PSF to simulate the effect of the aperture of a telescope is the most obvious operator, originally introduced in the context of the present study by \citetads{2012A&A...548A...5V}.  In the following we will assume a model grid of $10\times 10$~pixel$^2$. For each monochromatic map, the degradation operator would consist of a matrix where the PSF is centered in each diagonal element and spread along each row in vector form. Let's assume a PSF $P_{ji}$ where the subindexes $ji$ spam from $(-n_{psf}/2, n_{psf}/2)$, so that the center of the PSF is at $(j,i) = (0,0)$. Let's also assume that $\sum P_{ij} = 1$, so that the intensity level does not change when we apply the PSF to the data. 

Close to the edges of the images, we cannot include the full PSF because we have no information of the Sun outside our image, but we can assume that the image is mirrored in which case the PSF at can be scaled as follows:
\begin{equation}
	\bar{P}_{ji} = 
		\begin{cases}
		P_{ji} & $i=j=0$\\
		p_{corr} \times P_{ji} & i \neq 0 \ \& \ j \neq 0
		\end{cases}
\end{equation}
where $\bar{P}_{ji}$ is the modified PSF at the edges of the image,  $p_{corr} = (1 - P_{0,0}) / (P_{sum}-P_{0,0})$ and $P_{sum}$ is the sum of all elements of the PSF that are inside the FOV when the PSF is centered in our pixel.

\begin{figure}
\centering
\includegraphics[width=0.95\columnwidth]{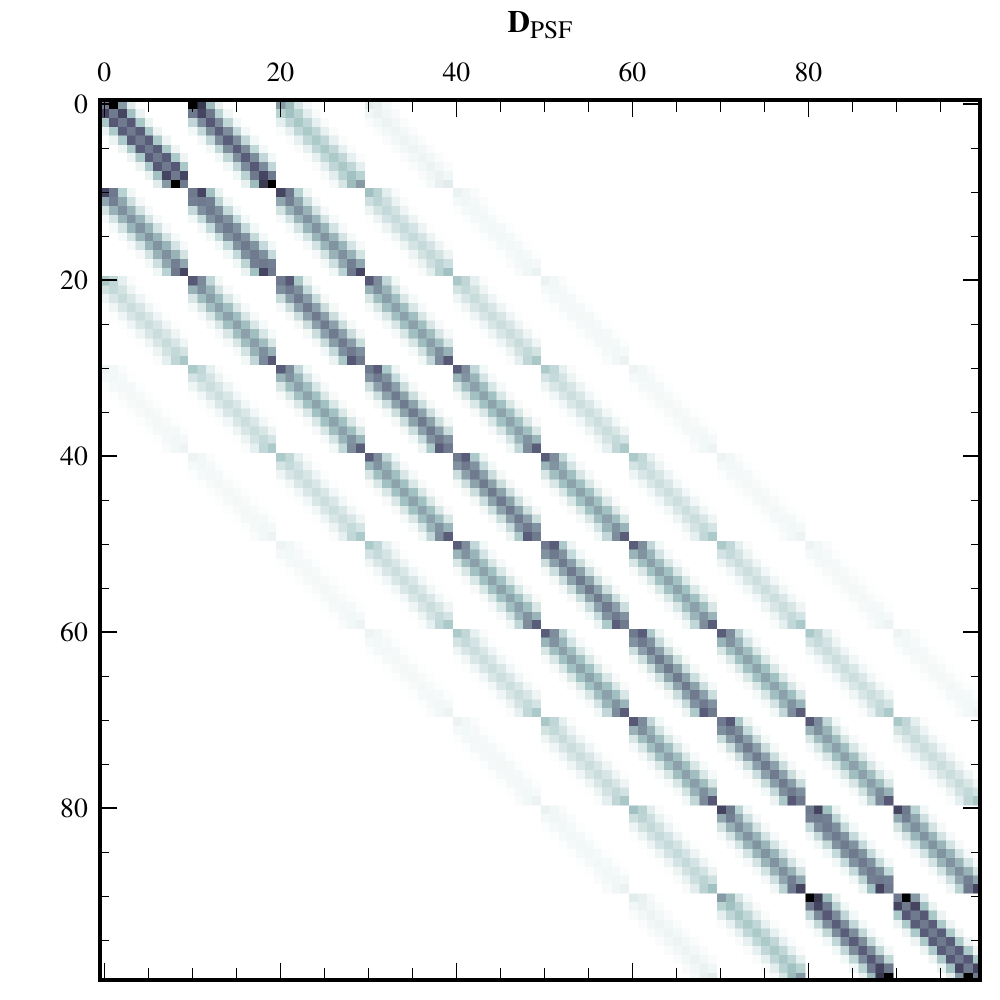}
\caption{Convolution operator using a 2D Gaussian PSF with $\mathrm{FWHM} = 2$~pixels and a FOV of $10\times 10$~pixels$^2$.}
\label{fig:oppsf}
\end{figure}

The convolution operator with a PSF is illustrated in Fig.~\ref{fig:oppsf}. Each row of the operator represents the degradation of one pixel, and therefore, it contains the contributions from the surrounding un-degraded pixels (given by the PSF).

\subsubsection{Rebinning/resampling operators}\label{sec:reb}
\begin{figure}
\centering
\includegraphics[width=0.95\columnwidth]{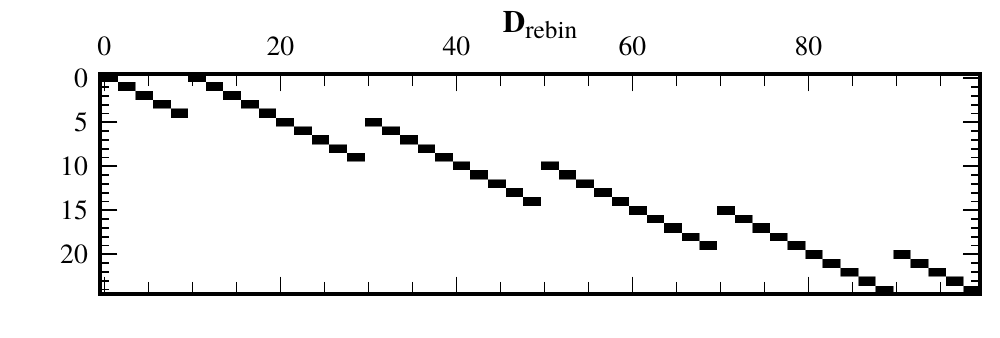}
\includegraphics[width=0.95\columnwidth]{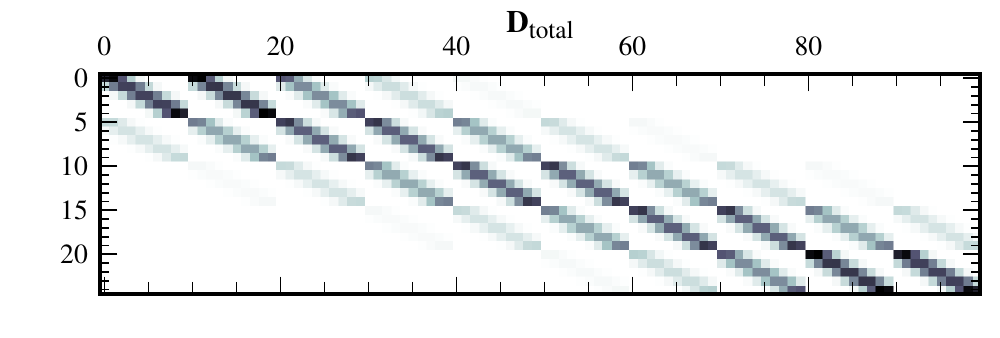}
\caption{\emph{Top:} representation of a $2\times 2$ rebinning operator for a $10\times 10$ pixel FOV. \emph{Bottom:} total operator resulting from the combination of a PSF degradation operator and a rebinning operator ($\mathbf{D}_{\mathrm{total}} = \mathbf{D}_{\mathrm{rebin}} \mathbf{D}_{\mathrm{PSF}}$).}
\label{fig:opreb}
\end{figure}

Rebining the data basically is similar to applying a square-box PSF, but it involves also a resampling of the data. The resampling operation is represented with a matrix with as many rows as resampled pixels, and as many columns as synthetic data points.
Therefore, the operator matrix is not square. Instead, we construct it with one row per rebinned pixel, with the corresponding square-box PSF centered in the required position over the FOV. The upper panel in Fig.~\ref{fig:opreb} illustrates the shape of a $2\times 2$ rebinning operator applied to a $10\times 10 $ pixels FOV. in this case, the number of rows should be $10/2 \times 10/2 = 25$.

The lower panel in Fig.~\ref{fig:opreb} illustrates the total combined operator $\mathbf{D}_{\mathrm{total}} = \mathbf{D}_{\mathrm{rebin}} \mathbf{D}_{\mathrm{PSF}}$, where we have combined a convolution with a PSF with a rebinning/resampling operator.

\begin{figure}
\centering
\includegraphics[width=0.95\columnwidth]{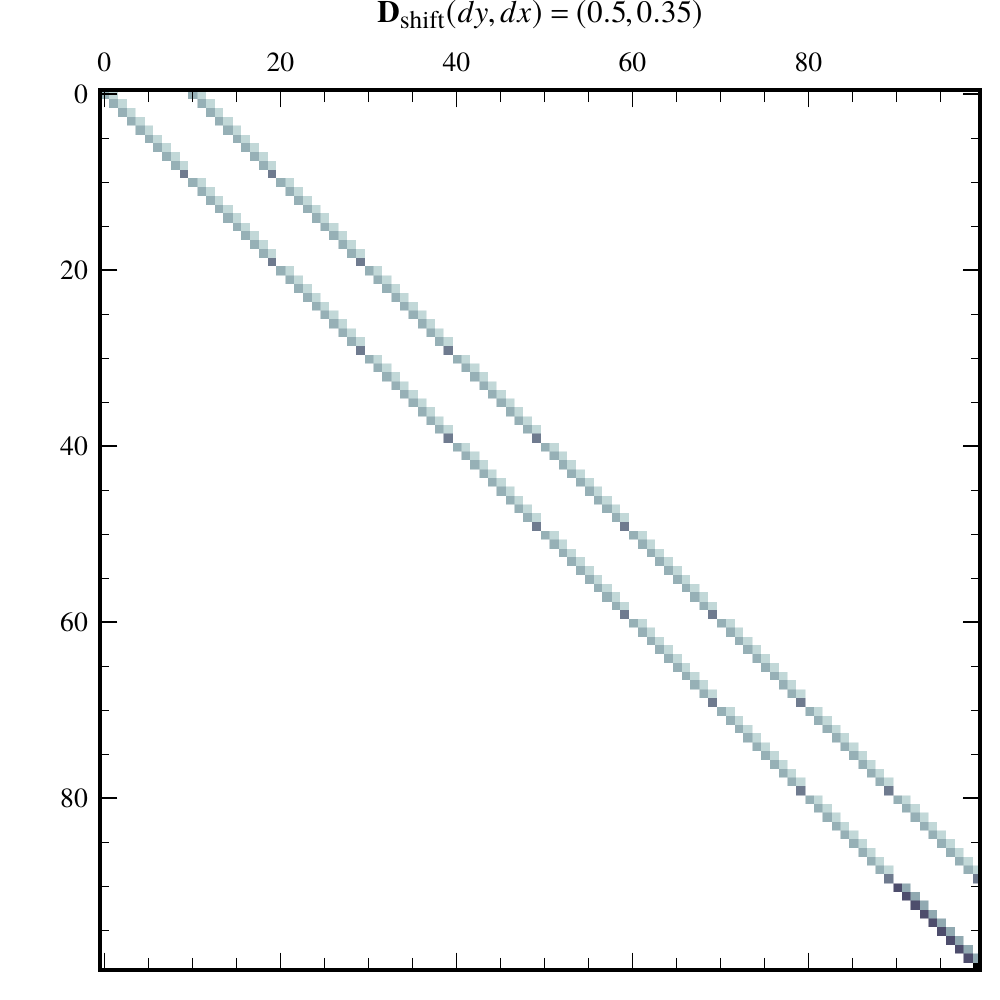}
\includegraphics[width=0.95\columnwidth]{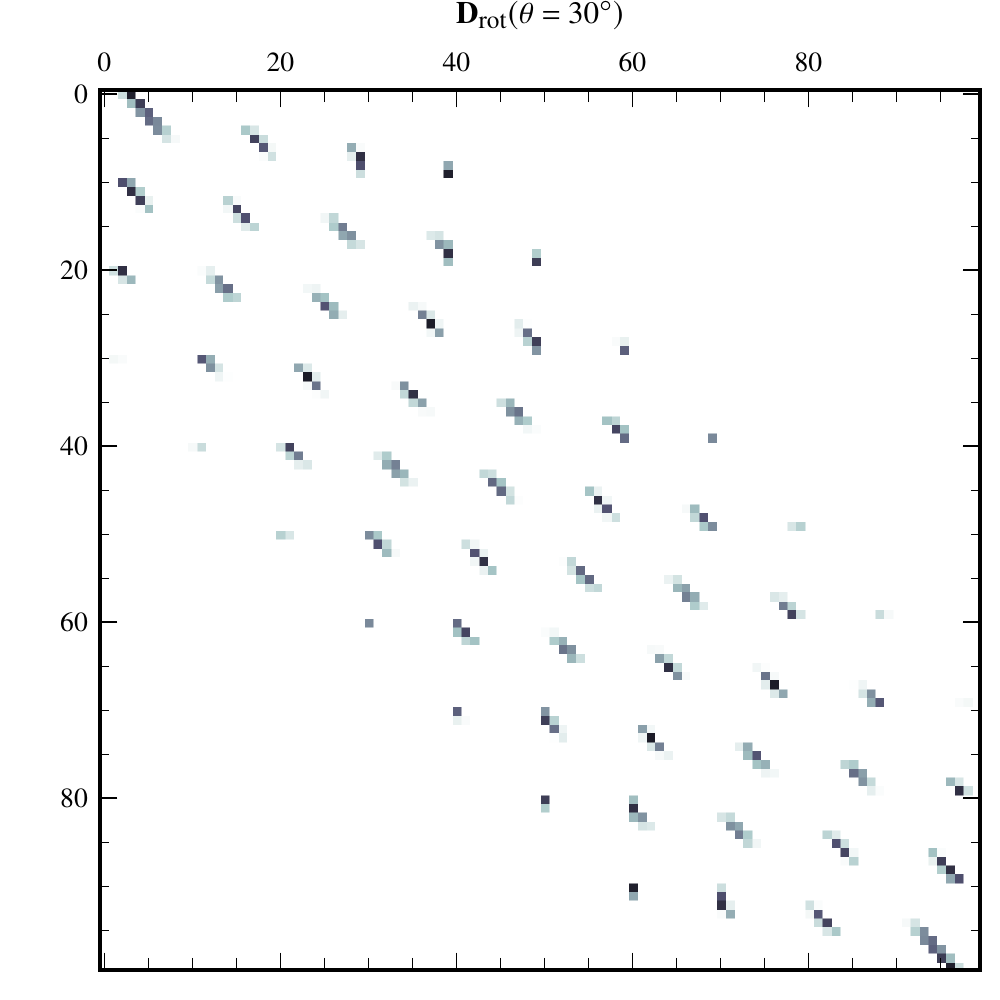}
\caption{Bilinear interpolation operators. \emph{Top}: constant image shift of $(dy,dx)=(0.50,0.35)$ over a FOV of $10\times 10$ pixels$^2$. \emph{Bottom}:  $30^\circ$ clockwise rotation for a FOV of $10\times 10$ pixels$^2$.}
\label{fig:opbil}
\end{figure}

\subsubsection{Interpolation operators}\label{sec:int}
In general, we might need to interpolate the synthetic spectra, which are noise-free, to adapt the data to a new grid or simply to apply rotations or shifts. When written in matrix form, an interpolation can be interpreted as a convolution with a PSF. In this case, the values of that PSF would depend on the type of interpolation (bilinear, bicubic) and the displacements. In any case, it can always be written as a linear combination of the values of the undegraded synthetic data. 

Bilinear interpolation is computed as a linear combination of four pixels. In Fig.~\ref{fig:opbil} we have illustrated bilinear interpolation operators for a constant shift over the entire image of $(dy,dx) = (0.50,0.35)$~pixels and for a clockwise image rotation of $\theta=30^\circ$. The balance between the contribution of the four pixels involved in each interpolations will of course change depending on the exact values of $(dy,dx)$, which in the case of a rotation will be different in each pixel. The rebinning $2\times 2$ operator has a very similar shape to a bilinear interpolation with a constant shift of $(dy,dx) = (0.5,0.5)$~pixels, where all four pixels contribute equally to the result, although the way the edges of the image are treated in the operator can be different and we have not considered a resampling of the data in this case as in rebinning.

\section{Numerical experiments}\label{sec:num}
We have implemented the inversion and degradation methods described in \S\ref{sec:gen} to create a proof-of-concept code, based on the Milne-Eddington model atmosphere (\citeads{1977SoPh...55...47A}). Radiative transfer calculations based on the Milne-Eddington atmosphere are very fast and the derivatives of the intensity relative to the model parameters can be obtained analytically (e.g. \citeads{2007A&A...462.1137O}). The latter has allowed for quick development and testing of the methods presented in this study.

Our aim is recreating a situation where two co-temporal datasets that provide complementary diagnostic information (acquired at different spatial resolutions) are used in one inversion that takes into account all instrumental effects consistently. In this case we will generate a high-resolution Stokes~$I$ observation in the \ion{Fe}{i}~6301~\AA\ line (telescope diameter $d_1=1$~m) that is combined with a lower resolution full-Stokes dataset acquired in the \ion{Fe}{i}~6302~\AA \ line ($d_2=0.5$~m). Therefore, most of the magnetic field information will be provided by the low resolution dataset, whereas both datasets will contribute to constrain the rest of the parameters of the model.


\subsection{3D rMHD simulation tests}
We have used a snapshot from a 3D rMHD simulation of an active region in the solar photosphere (\citeads{2012ApJ...750...62R}) that was performed with the MURAM code (\citeads{2005A&A...429..335V}). The simulation was performed with a horizontal resolution of $12$~km and $8$~km in the vertical direction, in a grid of dimensions $(n_x,n_y,n_z)=(4096,4096,256)$~pixels. The simulation snapshots that we used covers a physical domain of $(s_x,s_y,s_z)=(49,49,2)$~Mm.

\begin{figure}
\centering
\includegraphics[width=\columnwidth]{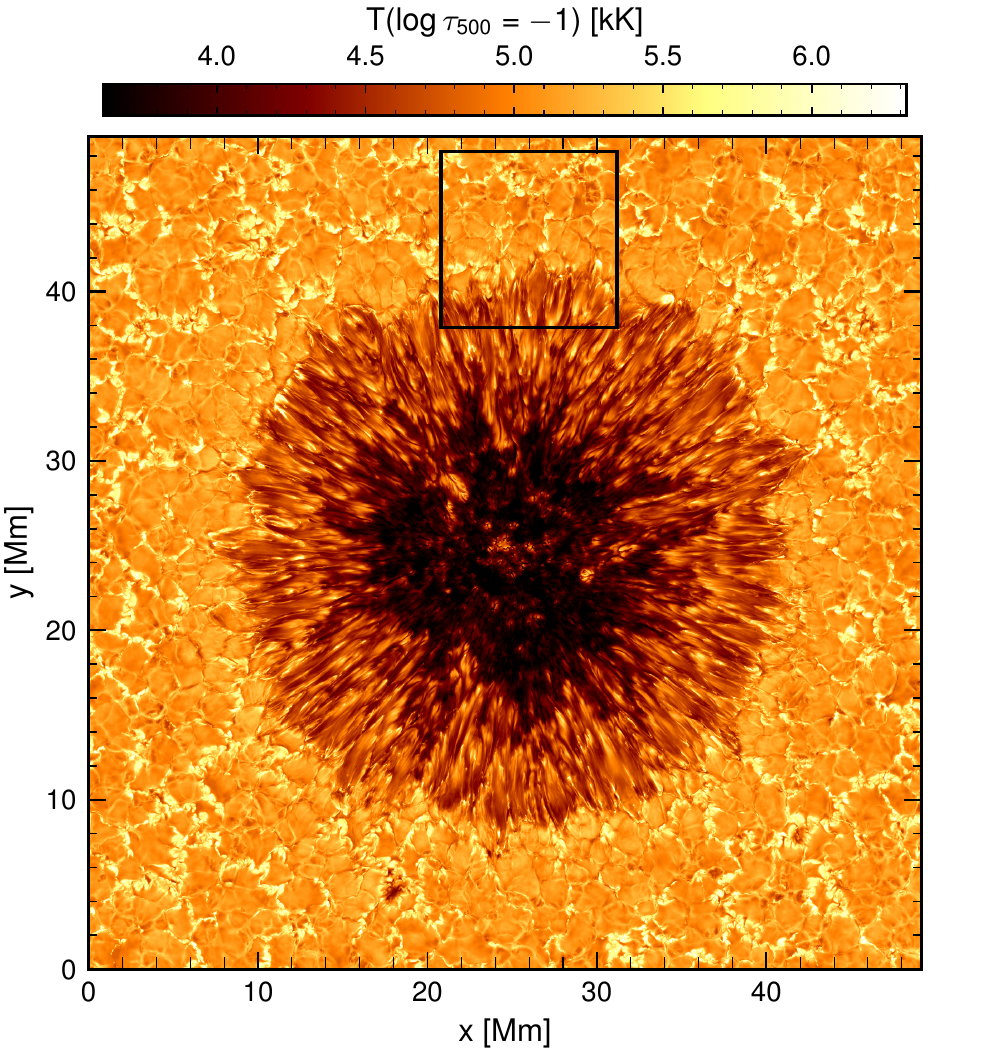}
\caption{Temperature of the 3D rMHD simulation (\citeads{2012ApJ...750...62R}) extracted where $\log\tau_{500} = -1$ in each column. The black box indicates the subfield of the simulation that we have used in our tests.}
\label{fig:modelall}
\end{figure}

In order to make the synthetic spectra symmetric so they better satisfy the limitations of the Milne-Eddington model atmosphere employed in the inversions, we have extracted the line-of-sight velocity and the three components of the magnetic field vector at the layer where $\tau_{500}=-1$ for each pixel and repeated that same value over the entire depth-stratification of each pixel. However, we have not imposed a linear source function in the synthetic observations, as it is assumed in the ME model atmosphere.

We have synthesized \ion{Fe}{i} spectra under the local-thermodynamical-equilibrium assumption using the STiC code (\citeads{2016ApJ...830L..30D}; \citeads{2019A&A...623A..74D}) over the dark box indicated in Figure~\ref{fig:modelall}. The spectra were computed in the photospheric \ion{Fe}{i}~6301\& 6302~\AA\ lines. The atomic data were extracted from the VALD-3 database (\citeads{2015PhyS...90e4005R}; \citeads{1995A&AS..112..525P}), including accurate van der Waals collisional cross-sections from collisions with neutral hydrogen (\citeads{2000A&AS..142..467B}).

\subsection{Synthetic observations}\label{sec:sobs}
\begin{figure}
\centering
\includegraphics[width=\columnwidth]{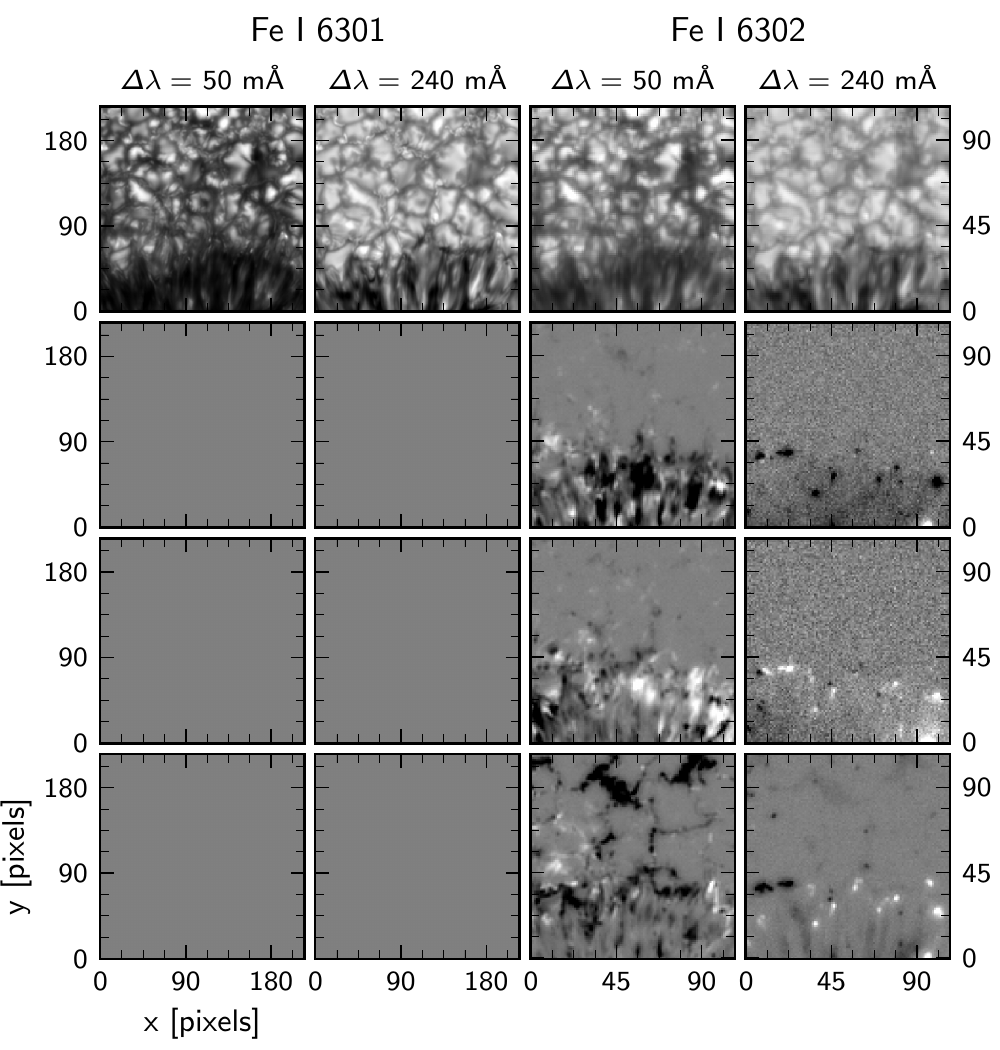}
\caption{Degraded synthetic data at two wavelengths of the \ion{Fe}{i}~6301~and of the 6302~\AA\ lines. The former has been degraded assuming a telescope with $d_1=1$~m and the latter with a telescope aperture $d_2=0.5$~m. Stokes~$Q$, $U$ and $V$ have been set to zero in the 6301 line dataset in order to perform numerical experiments. Each dataset has been sampled in a different spatial grid.}
\label{fig:data}
\end{figure}

\begin{figure}
\centering
\includegraphics[width=\columnwidth]{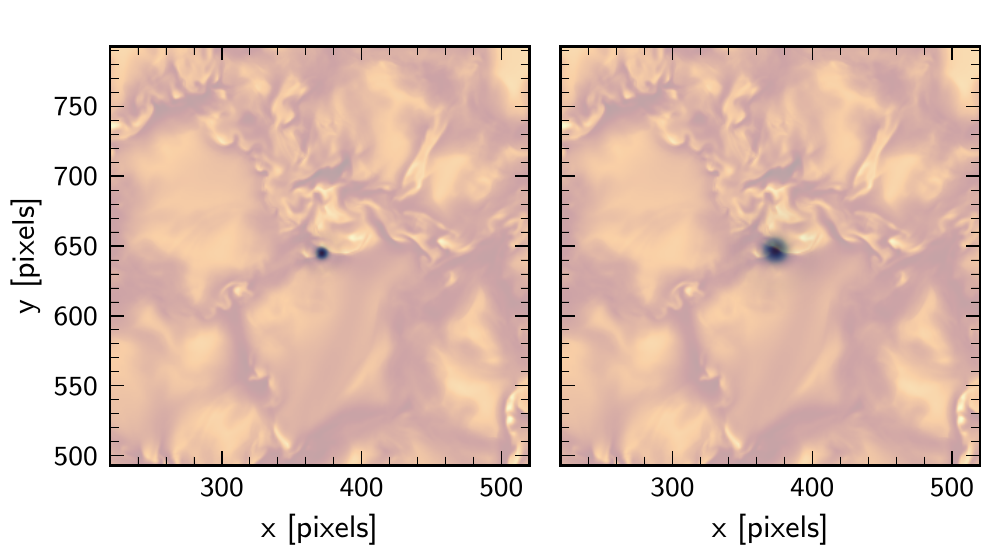}
\includegraphics[width=\columnwidth]{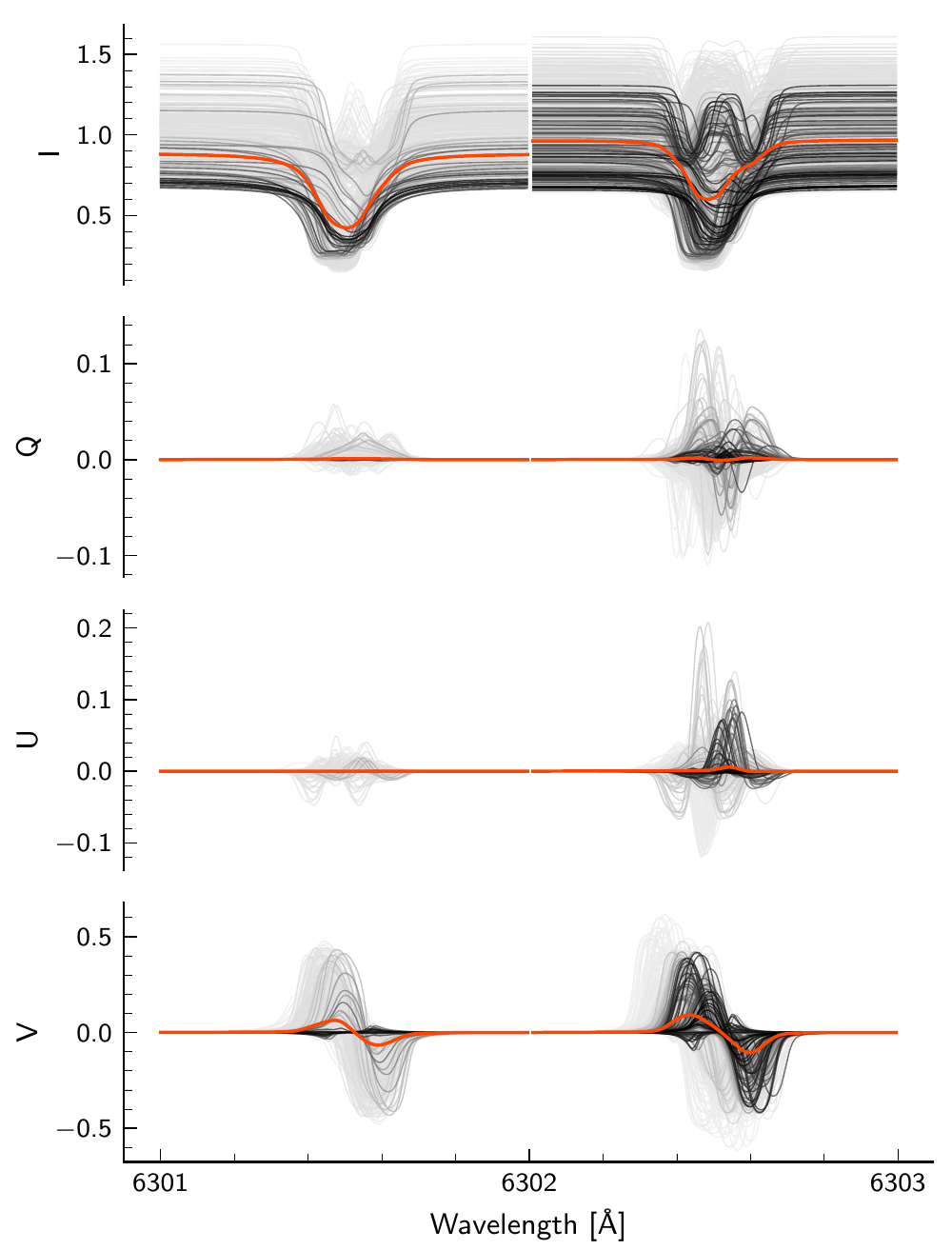}
\caption{Spatial degradation of the spectra computed from the rMHD simulation using two telescope apertures. \emph{Top:} Illustration of the extent of the total degradation operator (PSF + rebinning) shown as a dark shade centered in a crop of the FOV for the 6301 line (left) and the 6302 line (right). The intensity panels show the resulting degraded  spectra in red and the contribution from all profiles contained within the degradation operator. The color scaling of each profile corresponds to the weight of the operator in each pixel. }
\label{fig:datadeg}
\end{figure}

In order to test the methods proposed in \S\ref{sec:gen} we have degraded the data corresponding to each line using a different spatial PSF. 
\begin{table*}
\caption{Summary of the numerical experiments performed in this study.}             
\label{tab:sobs}      
\centering
\begin{tabular}{l | r |c c c c c c}          
\hline\hline                        
Description (model grid) & Remarks & $\lambda$ [\AA] & $D [m]$ & $\delta s$ [\arcsec] & rebin &grid size [pixels] & Stokes\\
\hline
Experiment 1& 1D inv &  6301.499 & 1.0 &  0.065& 1 &$216\times 216$ & $I$\\
($216\times 216$ pixels) & \ &\ &\ &\ &\ &\ \\
\hline                                   
Experiment 2& 1D inv &  6301.499 & 1.0 &  0.065& 1 &$216\times 216$ & $I$\\
($216\times 216$ pixels) & \ &6302.494 & 0.5 & 0.13 & 2 & $216\times 216$\tablefootmark{1} & $I,Q,U,V$\\
\hline
Experiment 3&  multi-resolution \& &  6301.499 & 1.0 &  0.065& 1 &$216\times 216$ & $I$\\\
($216\times 216$ pixels) & regularization &6302.494 & 0.5 & 0.13 & 2 & $108\times 108$ & $I,Q,U,V$\\
\hline
\end{tabular}
\tablefoot{$^{1}$The observation has been upsampled from a grid of $108\times 108$ to $216\times 216$ pixels using bilinear interpolation.}
\end{table*}
We have generated theoretical/idealized telescope PSFs by recreating a circular aperture with a central obscuration of $11\%$ of the total aperture area. We have not added a spider for the secondary, but we have considered an effective total defocus of less than $1/16$ of a wave. It is a similar telescope model as that considered by \citetads{2008A&A...484L..17D} for the Hinode SOT instrument, but we have ignored the effect of the spider and changed the aperture size. After degradation, the two datasets have been critically resampled according to Nyquist's theorem by rebinning. In the case of the \ion{Fe}{i}~6301~\AA\ line, we have created a telescope PSF assuming an aperture $d_1 = 1$~m and a rebinning factor $\times 4$ which corresponds almost exactly to the diffraction limit of a $1$~m telescope, whereas for the 6302~\AA\ line we have used an aperture $d_2=0.5$~m and a rebinning factor $\times 8$. The total degradation operator is therefore similar to that shown in the lower panel of Figure~\ref{fig:opreb}. We have set to zero Stokes~$Q$, $U$ and $V$ of the 6301 dataset. Figure~\ref{fig:data} illustrates monochromatic images from the resulting datasets at two different wavelengths in each of the spectral lines.

The effects of the total degradation operator for each dataset (PSF and rebinning) are illustrated in Figure~\ref{fig:datadeg}. The extension of the PSF is depicted in the upper panels for each of the lines. The observed spectra (red) is a weighted average of many smaller scale spectra enclosed within the extent of the PSF, which are also plotted and colored in grey-scale according to the weight of the PSF. Note that all individual profiles are strictly symmetric because we have removed all vertical velocity gradients from the MHD simulation, but the resulting degraded spectra are highly asymmetric as a result of the spatial average. 

We have added Gaussian noise to the synthetic data with $\sigma_{6301}(I,Q,U,V)=(10^{-3},10^{30},10^{30},10^{30})$ and  $\sigma_{6302}(I,Q,U,V)=(10^{-3},10^{-3},10^{-3},10^{-3})$ relative to the mean continuum intensity over the FOV. 
We have assumed a Gaussian spectral degradation of $\mathrm{FWHM} = 20$~m\AA.

\subsection{Experiments setup}
We have recreated three situations that should allow to visualize the power of our new method:
\begin{itemize}
\item Experiment 1 is a traditional 1D inversion using only the highest resolution data, which only contains the intensity spectra. This inversion is a benchmark to assess how much detail can be retrieved from the highest resolution dataset that we have considered.
\item Experiment 2 corresponds to a case where we have resampled the low resolution data acquired in the 6302~\AA \ line to match the spatial grid of the 6301~\AA\ data using bilinear interpolation, and we perform a traditional 1D inversion including both lines. This case is representative of a situation where data acquired at different spatial resolution are combined in one inversion without accounting for spatial degradation.
\item Experiment 3 corresponds to a multi-resolution global inversion where the synthetic data are degraded and resampled to match the observations, which are given in different spatial grids.
\end{itemize}
Some of the properties of these experiments are summarized in Table~\ref{tab:sobs}. All experiments were initialized using the same constant model over the entire FOV, which is given in the same grid as the high resolution data ($0.065\arcsec/\mathrm{pixel}$).

In the case of the 1D inversions, we performed nine additional inversions of each pixel starting from different initial conditions, where we kept the best fit of all of them. Afterwards, the result was spatially smoothed and one final inversion was performed in each pixel from that smoothed result.

The multi-resolution inversion was performed in two cycles. In the first cycle we used a largely overestimated ($\times 50$) weight for the regularization functions. The latter allows to retrieve in very few iterations the large scale properties of the model (usually in less than 10 iterations). Then we reduced the regularization weight to a correct value (discussed below) and re-started the inversion process in order to fit the small scale details of the model. In our tests, the second part of the inversion was also finished in less than 10 iterations. No smoothing was applied to the output model between the two cycles.

\section{Results}\label{sec:res}
\begin{figure*}
\centering
\includegraphics[trim={0 0.8cm 0 0}, clip,width=\textwidth]{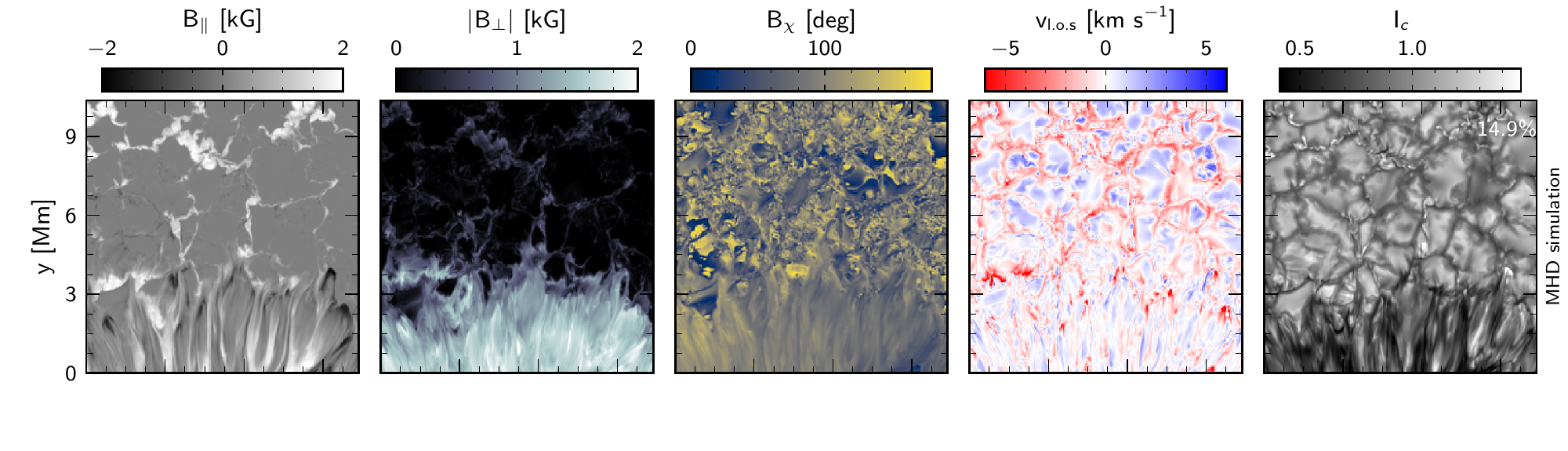}
\includegraphics[trim={0 0.8cm 0 1.2cm}, clip,width=\textwidth]{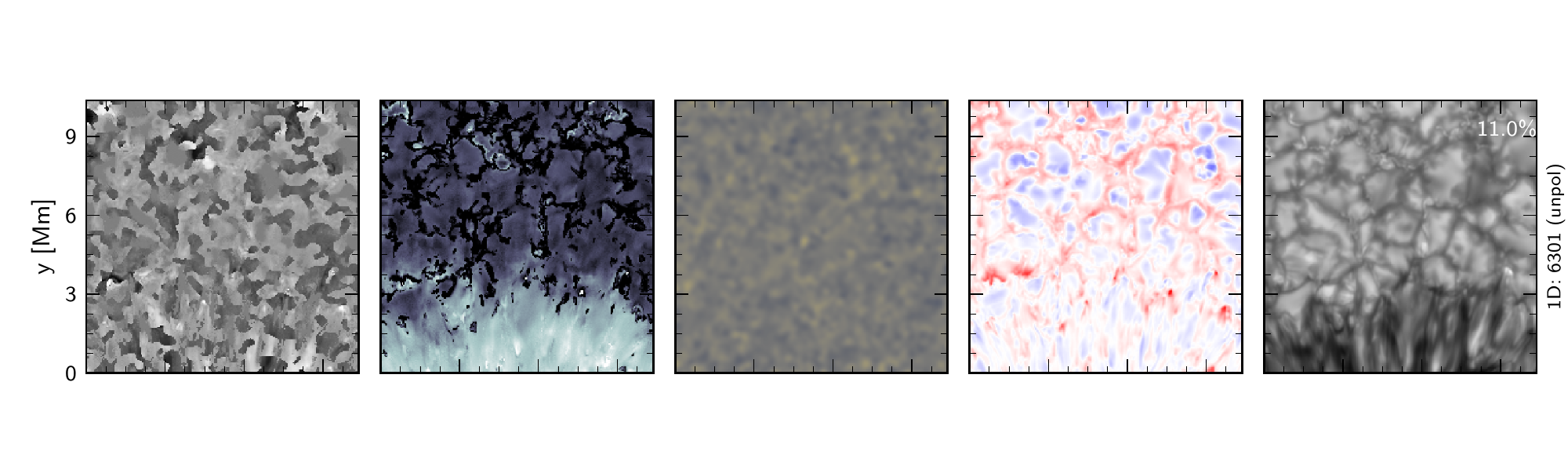}
\includegraphics[trim={0 0.8cm 0 1.2cm}, clip,width=\textwidth]{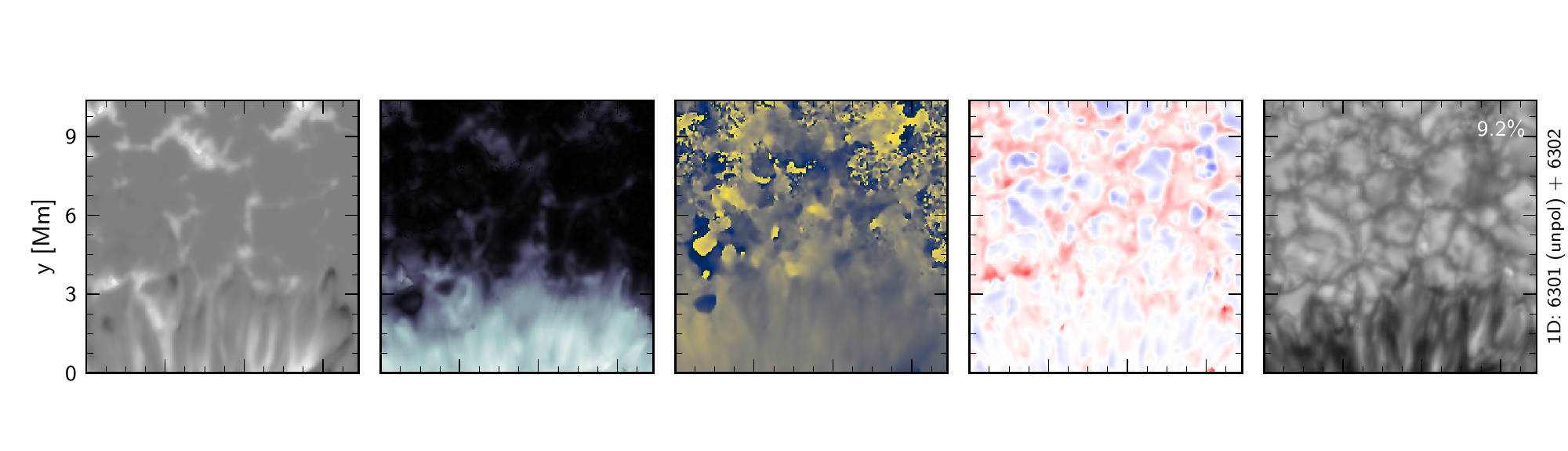}
\includegraphics[trim={0 0.cm 0 1.2cm}, clip,width=\textwidth]{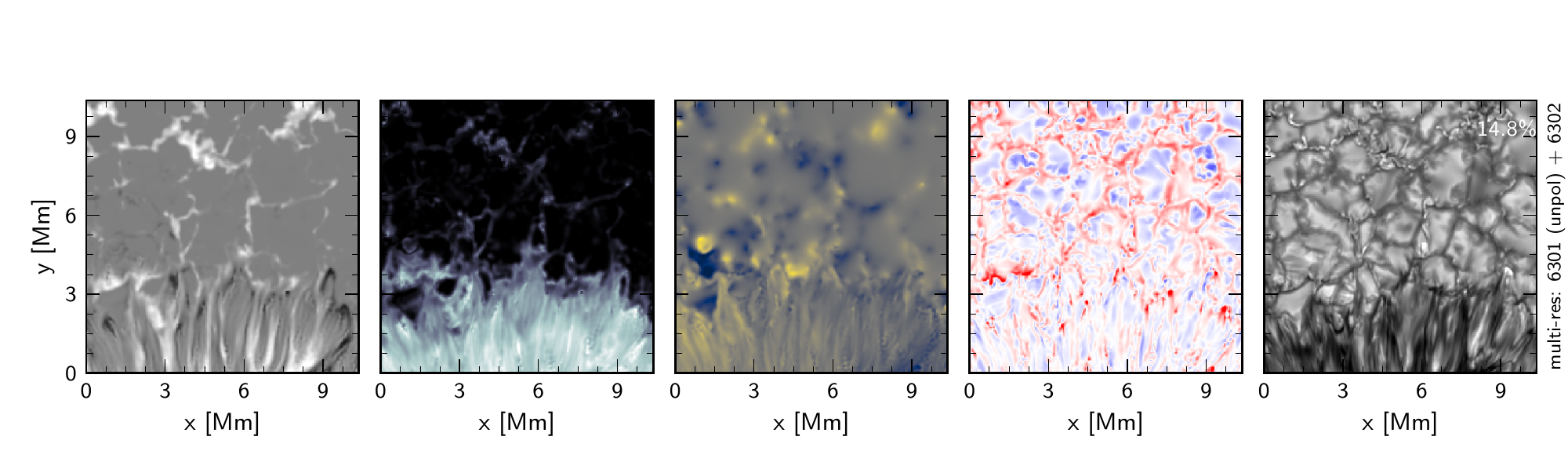}
\caption{From left to right: longitudinal component of the magnetic field, absolute value of transverse component of the magnetic field, magnetic field azimuth, line-of-sight velocity and continuum intensity. From top to bottom: original rMHD simulation (which has been resampled to the grid of the highest resolution dataset), inversion results from experiment 1, results from experiment 2 and results from experiment 3 (see table~\ref{tab:sobs}). The granulation continuum intensity contrast is marked in the rightmost panel. The continuum intensity in the lower three rows in computed as $I_c = S_0+S_1$.}
\label{fig:maps}
\end{figure*}
\begin{figure*}
\centering
\includegraphics[trim={0 0.685cm 0 0.8cm}, clip,width=\textwidth]{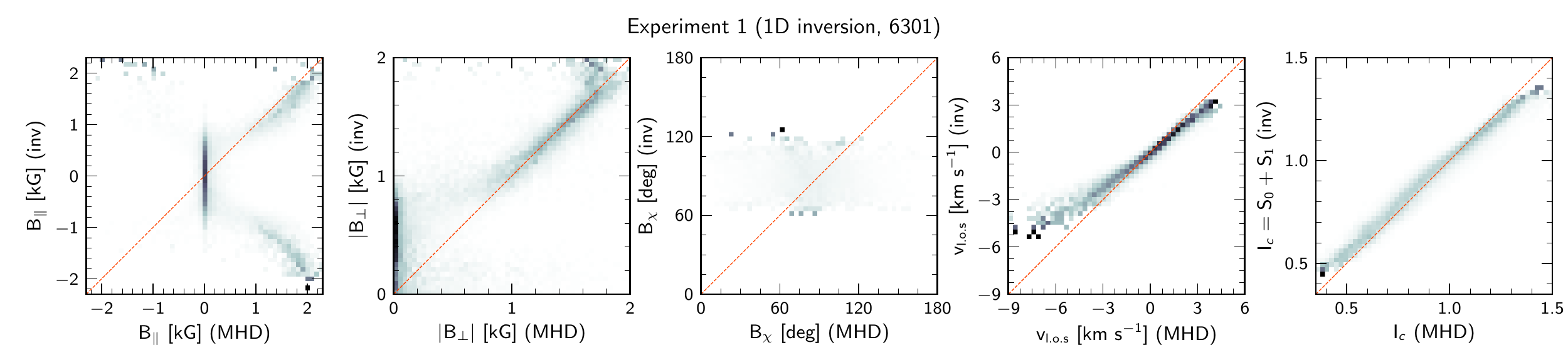}
\includegraphics[trim={0 0.685cm 0 0.8cm}, clip,width=\textwidth]{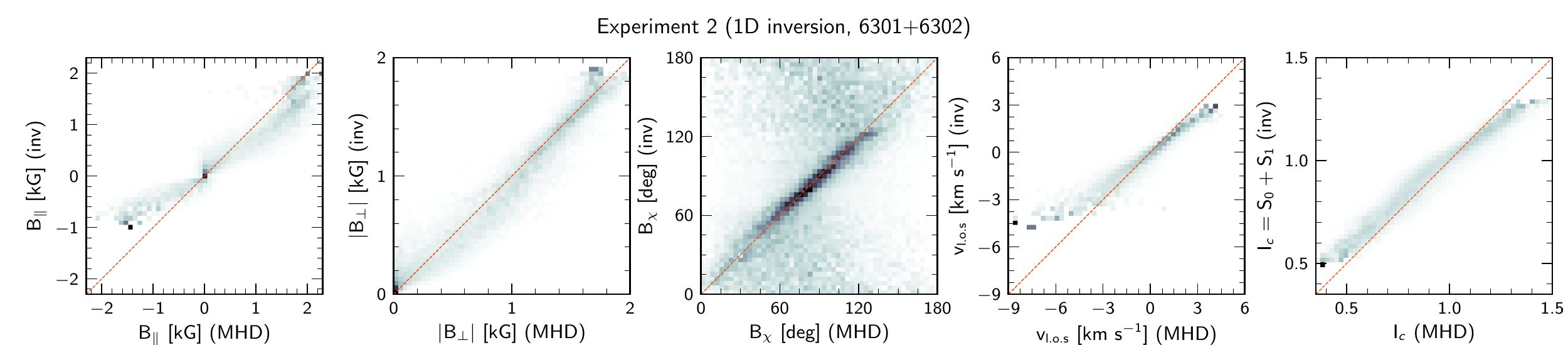}
\includegraphics[trim={0 0cm 0 0.8cm}, clip,width=\textwidth]{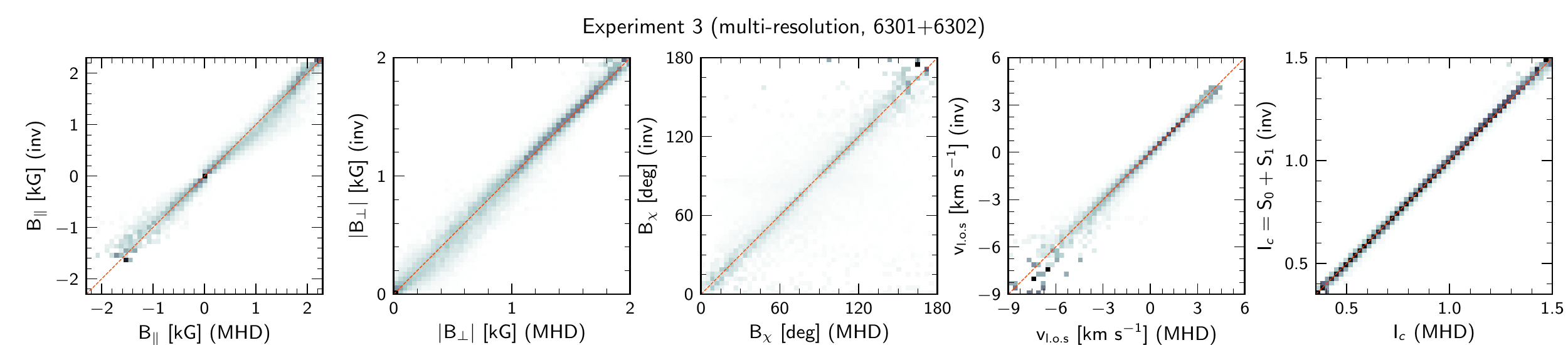}
\caption{Density maps comparing the inversion results from Fig.~\ref{fig:maps} to the parameters of the rMHD model. The panels are arranged in the same order as in the former figure. Each row has been normalized by the integral of the histogram along that row, to better illustrate the spread around each value. }
\label{fig:scat}
\end{figure*}
The results from the three experiments are summarized in Figures~\ref{fig:maps} and \ref{fig:scat}, where the inversion results are compared with the quantities from the original rMHD simulation. The snapshot is computed in a spatial grid with cell size $ds=12$~km and it contains more fine structure than an ideal 1-m aperture telescope can retrieve. Therefore, for this comparison we have applied a Fourier frequency cut-off corresponding to the limit frequency of a 1-m telescope and then we have resampled the model using $4\times 4$ rebinning, which yields resolution elements of $0.065\arcsec/\mathrm{pixel}$. This way, we have limited the amount of detail of the input model to that allowed by a 1-m telescope and we have sampled the model in the same grid as the 6301 line dataset.

We have included in this comparison the three components of the magnetic field vector, the line-of-sight velocity map and the continuum intensity. The continuum intensity that is displayed for the inversions corresponds to the sum of the source function terms derived from the Milne-Eddington model ($I_c = S_0 + S_1$). Since the input spectra were computed under the assumption of local thermodynamical equilibrium, the source function terms relate to the thermal stratification of the MHD atmosphere, and the continuum intensity is a proxy of how well that part is reconstructed. We have indicated the RMS contrast of the continuum intensity computed on the upper $2/3$ of the FOV. The intensity RMS contrast in the reference MHD simulation is $14.9\%$.
\begin{figure}
\centering
\includegraphics[width=\columnwidth]{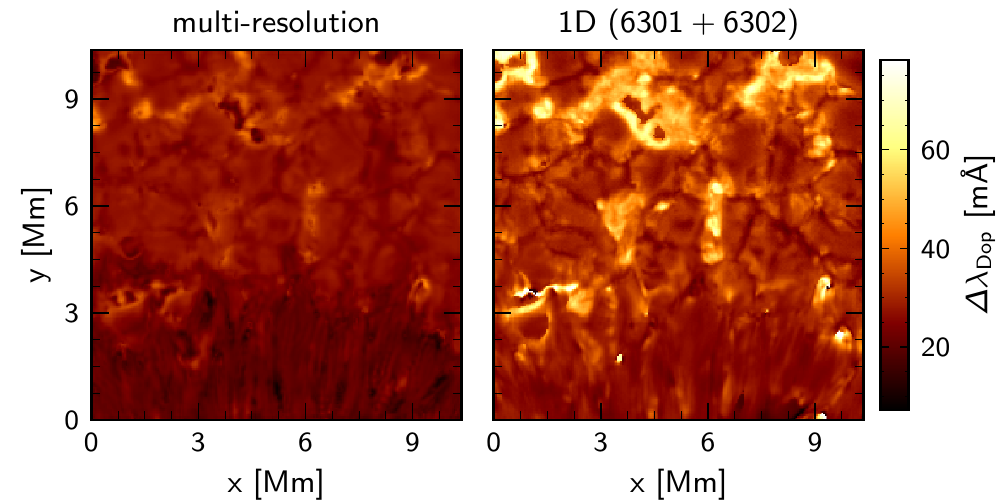}
\caption{Doppler-width inferred from the multi-resolution inversion (left) and from the 1D inversion including both spectral lines.}
\label{fig:mic}
\end{figure}

The results from experiment 1 show that the magnetic field is largely unconstrained by using Stokes $I$ only, although some information of the magnetic field strength could be retrieved in the penumbra included in the FOV (e.g., \citeads{1990ApJ...355..329L}). The decomposition that we have chosen seems to accumulate most of the magnetic field signal in $|B_\bot |$, but that is purely due to the geometry of the magnetic field and the Zeeman imprint in Stokes $I$. The effect of the telescope PSF degradation is quite apparent in the velocity and continuum intensity maps, where the contrast is significantly lower than in the MHD model. The scatter plots in Figure~\ref{fig:scat} also shows that the amplitudes of the velocity map are systematically underestimated in upflowing and downflowing regions due to the blurring effect of the 1~m aperture that we have considered in this experiment. This effect is not as obvious in the continuum intensity scatter plot, but still present. In this case, the recovered continuum intensity contrast is $11.0\%$.

In experiment 2 we have added a full-Stokes dataset acquired at lower resolution. A priori we would expect to largely improve the recovery of the magnetic field vector, at the cost of degrading the resolution and contrast of the velocity and continuum intensity maps, and that is to a large extend what the 1D inversion recovers. The three components of the magnetic field vector now resemble the structures present in the MHD model, but cancellation effects between opposite polarities start to affect the recovery of Stokes~$V$, as illustrated in the scatter plot of $B_{\parallel}$ for this experiment. The azimuth of the magnetic field is mostly recovered in the penumbra where the magnetic field is strong, and largely random outside the spot as there are large parts of the FOV without sufficient signal to compute the magnetic field vector. In this case, the continuum intensity contrast is decreased to $9.2\%$ as a result of adding the lower resolution data but also due to the discrepancies between both datasets.

Experiment 3 properly addresses instrumental degradation in each of the datasets: the resulting maps are naturally corrected for the telescopes PSFs and sampled to the correct grid of each observation before computing $\chi^2$. All derived quantities follow the one-to-one line indicated in the scatter plots, indicating that the amplitudes and contrast of all quantities are correct. In the maps of the magnetic field vector it is also clear that the resolution of the magnetic field cannot be constrained at the same resolution as the velocity and continuum intensity maps, as most (but not all) of the information is derived from the polarimetric low resolution dataset. However, it is remarkable how each of the datasets have helped to set constraints on different variables of the model atmosphere without negatively affecting the result and at the resolution that is present in each dataset. We recover the original continuum intensity contrast almost entirely, with an RMS value of $14.8\%$. The slightly lower value could be due to the fact that the source function in the original data is not forced to have a linear dependence with the continuum optical depth ($\tau_c$), which is an intrinsic assumption of the Milne-Eddington atmosphere. Regularization can certainly have a smoothing effect in the result if its weight is not properly adjusted, but we chose a final value for the second cycle that allowed to converge the problem without showing artifacts from the deconvolution. 

Degraded profiles are systematically broader than un-degraded ones because they result from a weighted average over profiles with different Doppler shifts and thermal properties. Figure~\ref{fig:mic} illustrates how the 1D inversion compensates this effect by increasing the Doppler-width of the lines close to regions with strong horizontal gradients in velocity or in magnetic patches. In comparison, the mutli-resolution inversion retrieves lower values over the entire FOV. 

\subsection{The effects of spatial regularization}
So far we have not discussed in detail the effects of spatial regularization. One of the main motivations to include these terms in the equations relates to the study of \citetads{2012A&A...548A...5V}. They reported that their algorithm required more human intervention than an traditional 1D inversion because the solution got stuck in local minima easily. The latter is probably due to the small-scale structure that the algorithm can try to introduce within the size of the PSF in the model atmosphere. When the solution is very far from the global minimum, the code can behave erratically, introducing small scale artifacts.

When properly used, regularization can help to address those issues in a very effective way. By choosing a very large and clearly overstimated value of regularization, we can force the algorithm to fit only large spatial scale features in the data during the first iterations. When $\chi^2$ has decreased by a few orders of magnitude, we can reset the spatial regularization weight to an appropriate value that allows to converge the model without affecting the overall quality of the fits. That is the criteria to choose the regularization weight. The latter can be very easily calibrated with a few test runs over a very small subpatch of the data.

In this application, regularization also had an unexpected secondary effect. We used an iterative BiCGStab method to solve the sparse linear system that provides the correction to the model in each iteration. When regularization was switched off and the diagonal damping parameter was small $(\lambda < 0.1)$, the linear solver would not properly converge to an accurate solution rapidly. When regularization was switched on, even if it was underestimated, machine precision accurate results could be obtained rapidly. This behaviour can be somewhat explained by looking at the structure of the regularization Hessian matrix in Fig.~\ref{fig:tik}. The latter contribution drives the solution in such a way that relatively smooth corrections are preferred and noisy solution that could be compatible with the data are discarded.

\subsection{The effects of noise}
\begin{figure}
\centering
\includegraphics[width=\columnwidth]{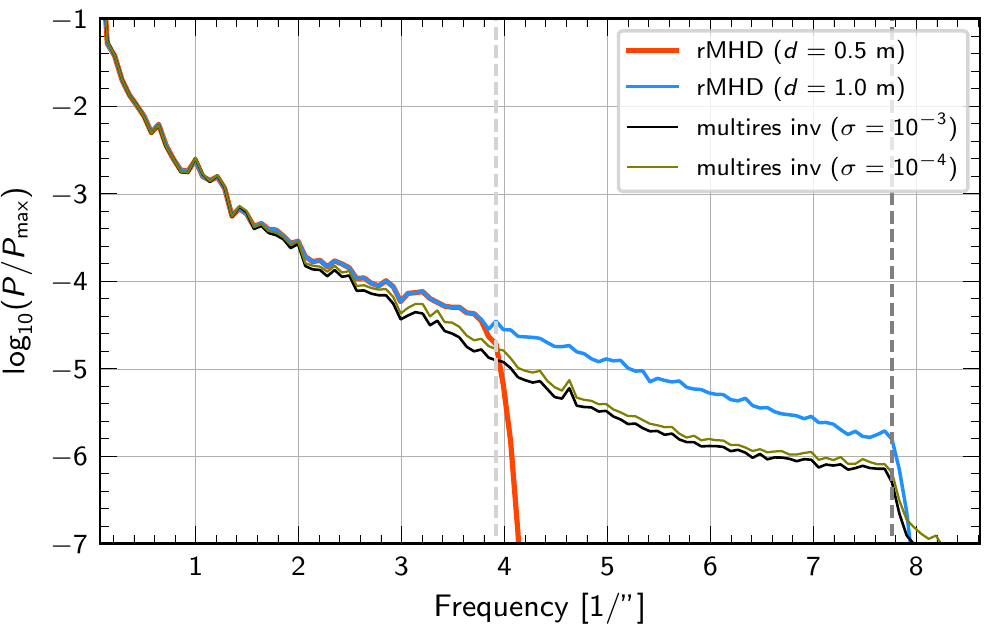}
\caption{Azimutal integral of 2D power spectra computed from $B_\parallel$ maps: from the multires inversion (black), from the rMHD simulation with a frequency cut-off filter corresponding to $d=1$~m (blue) and $d=0.5$~m (red). The theoretical cut-off limits of $0.5$~m and $1.0$~m aperture telescopes ($\lambda/d$) are indicated with a light-grey and dark-grey vertical lines. The green curve illustrates the power spectra of a multiresolution inversion performed with less noisy data ($\sigma=10^{-4}$).}
\label{fig:pow}
\end{figure}
The aperture size of any given telescope sets the maximum spatial resolution of the data that we can acquire with it. However, the presence of noise can lower that theoretical limit because the highest frequencies can become indistinguishable from noise. Figure~\ref{fig:pow} shows azimuthally-averaged power spectra computed from the $B_\parallel$  maps of the multiresolution inversion and from maps from the rMHD simulation where we have applied a frequency cut-off filter corresponding to $d=1$~m and $d=0.5$~m. The plots show that the multiresolution reconstruction has power beyond the cut-off limit of a $0.5$~m aperture. One could wonder how is it possible to retrieve magnetic field information with power above the cut-off frequency of the polarimetric data. However the higher spatial resolution dataset ($6301$) also contributes constraining the magnetic field via the Zeeman splitting of the Stokes~$I$ profile. We have tried to identify why there is lower power in high frequencies compared to the rMHD model and which magnetic structures in the FOV are most affected. We have calculated similar power-spectra in a patch within the umbra and in a small patch of the FOV containing weak small-scale fields. From those power spectra we can conclude that the discrepancy between the blue curve and the black curve originates mostly (but not all) in relatively weak fields concentrated in very small spatial scales. Those fields are not sufficiently strong to produce Zeeman splitting in Stokes~$I$, so the higher resolution dataset does not help constraining them. Part of this effect is also visible in the magnetic field maps displayed in Fig.~\ref{fig:maps}.

We have repeated the multi-resolution inversion with a dataset that had a much lower noise level with $\sigma_{6301}(I,Q,U,V)=(10^{-4},10^{30},10^{30},10^{30})$ and $\sigma_{6302}(I,Q,U,V)=10^{-4}$. The aim of this experiment is to assess whether the assumed noise level in our experiments has limited the amount of detail that could be retrieved. The reconstruction of the dataset with $\sigma=10^{-4}$ is able to recover slightly more high-frequency power than the noisier case, but the increase in this case is small (see Fig.~\ref{fig:pow}). We believe that there is no major difference between these two reconstructions because for both assumed noise levels, the imprint of Zeeman splitting in the Stokes~$Q,U,V$ signals is above the noise for most structures present in the FOV, which have large magnetic field strength.

\section{Discussion and conclusions}\label{sec:dcon}
We have developed a new method that allows for the simultaneous inversion of datasets acquired with different spatial resolution. We have generalized the Levenberg-Marquardt algorithm to deal with any form of degradation or transformation that need to be applied to the data in order to recreate the observations. These transformations are applied in the form of a chain of linear operators that modify the synthetic spectra during the inversion process. We have shown that our method extracts information from two datasets where each of them set constraints in the parameters of a model atmosphere at different scales. 

The inclusion of an extended telescope PSF is computationally very challenging, as it was originally shown by \citetads{2012A&A...548A...5V}. It is largely a matter of storage in memory and resolution of a very large linear system. The methods described in this paper also allow for an approximate approach that simplifies these computations enormously. The input data can be explicitly deconvolved using the telescope PSF as a pre-processing step. The only remaining part would be the resampling operations, which are very simple to implement and they require less memory. Even if no spatial transformations are required to transform the synthetic data, we can keep the inversion problem global by keeping the regularization terms in the problem, which couples the blocks shown in Fig.~\ref{fig:blkdiag}. The gains of having a global problem are mostly related to removing degeneracies from the solution of the inversion because the solution of one pixel has to be consistent with the model derived in its surroundings.

In this study we have assumed that all pixel straylight originates from the effect of the telescope aperture. But, datasets acquired at ground-based facilities are often processed with image reconstruction techniques like multi-object-multi-frame-blind-deconvolution (MOMFBD, \citeads{2002SPIE.4792..146L}; \citeads{2005SoPh..228..191V}) that compensate for this effect. However, the present method could still be used to compensate residual high-order aberrations induced by seeing (\citeads{2010A&A...521A..68S}), which also lower the contrast of the data. 

The present method is particularly relevant in the context of chromospheric research, where the very restricted selection of spectral lines that sample the chromosphere usually has forced our community to combine datasets acquired at different spatial resolution. That resolution discrepancy can originate from the combination of data acquired in different facilities and/or from the combination of data acquired with the same telescope at very different spectral regions. It has also become clear in early development studies of large-aperture telescopes that, at the diffraction limit, the photon count does not increase and therefore high-sensitivity polarimetric data will be acquired at lower spatial resolution. The resolution gap between ground-based and space-borne facilities will increase with the arrival of the new 4-m class telescopes (DKIST and EST) and the only way to properly combine as much data as is available is using global methods like the ones described in this study. 

Different instruments and facilities operate at different temporal cadences. We have not considered the effects of small temporal inconsistencies that could appear when the datasets included in the inversion are not strictly co-temporal. Future studies should aim at evaluating and proposing strategies for diminishing these effects (e.g., time interpolation).

In the future, we will include the development presented in this study in the STiC code to properly address chromospheric problems.

\begin{acknowledgements}
I am grateful to G. Scharmer, M. L\"ofdahl, M. van Noort, S. Danilovic, J. Leenaarts and M. Collados Vera for illuminating discussions about the content of this manuscript.
I am most grateful to M. Rempel for making his 3D rMHD simulations publicly available. JdlCR is supported by grants from the Swedish Research Council (2015-03994), the Swedish National Space Board (128/15) and the Swedish Civil Contingencies Agency (MSB). This project has received funding from the European Research Council (ERC) under the European Union's Horizon 2020 research and innovation programme (SUNMAG, grant agreement 759548). 
 Computations were performed on resources provided by the Swedish National
 Infrastructure for Computing (SNIC) at the PDC Centre for High Performance Computing (Beskow, PDC-HPC)
 at the Royal Institute of Technology in Stockholm as well as recourses at the National Supercomputer Centre (Tetralith, NSC) at Link\"oping University.
This study has been discussed
within the activities of team 399 'Studying magnetic-field-regulated heating in the solar chromosphere' at the International Space Science Institute (ISSI) in Switzerland.
The Swedish 1-m Solar Telescope is operated on the island of La Palma by the Institute for Solar Physics of Stockholm University in the Spanish Observatorio del Roque de los Muchachos of the Instituto de Astrof\'isica de Canarias.
The Institute for Solar Physics is supported by a grant for research infrastructures of national importance from the Swedish Research Council (registration number 2017-00625). 
 \end{acknowledgements}

\bibliographystyle{aa} 
\bibliography{references}

\begin{thebibliography}{43}
\expandafter\ifx\csname natexlab\endcsname\relax\def\natexlab#1{#1}\fi

\bibitem[{{Asensio Ramos} \& {de la Cruz
  Rodr{\'{\i}}guez}(2015)}]{2015A&A...577A.140A}
{Asensio Ramos}, A. \& {de la Cruz Rodr{\'{\i}}guez}, J. 2015, \aap, 577, A140

\bibitem[{{Auer} {et~al.}(1977){Auer}, {Heasley}, \&
  {House}}]{1977SoPh...55...47A}
{Auer}, L.~H., {Heasley}, J.~N., \& {House}, L.~L. 1977, \solphys, 55, 47

\bibitem[{{Barklem} {et~al.}(2000){Barklem}, {Piskunov}, \&
  {O'Mara}}]{2000A&AS..142..467B}
{Barklem}, P.~S., {Piskunov}, N., \& {O'Mara}, B.~J. 2000, \aaps, 142, 467

\bibitem[{{Borrero} {et~al.}(2017){Borrero}, {Franz}, {Schlichenmaier},
  {Collados}, \& {Asensio Ramos}}]{2017A&A...601L...8B}
{Borrero}, J.~M., {Franz}, M., {Schlichenmaier}, R., {Collados}, M., \&
  {Asensio Ramos}, A. 2017, \aap, 601, L8

\bibitem[{{Buehler} {et~al.}(2015){Buehler}, {Lagg}, {Solanki}, \& {van
  Noort}}]{2015A&A...576A..27B}
{Buehler}, D., {Lagg}, A., {Solanki}, S.~K., \& {van Noort}, M. 2015, \aap,
  576, A27

\bibitem[{{Buehler} {et~al.}(2019){Buehler}, {Lagg}, {van Noort}, \&
  {Solanki}}]{2019arXiv190807464B}
{Buehler}, D., {Lagg}, A., {van Noort}, M., \& {Solanki}, S.~K. 2019, arXiv
  e-prints, arXiv:1908.07464

\bibitem[{{da Silva Santos} {et~al.}(2018){da Silva Santos}, {de la Cruz
  Rodr{\'\i}guez}, \& {Leenaarts}}]{2018A&A...620A.124D}
{da Silva Santos}, J.~M., {de la Cruz Rodr{\'\i}guez}, J., \& {Leenaarts}, J.
  2018, \aap, 620, A124

\bibitem[{{Danilovic} {et~al.}(2008){Danilovic}, {Gandorfer}, {Lagg},
  {Sch{\"u}ssler}, {Solanki}, {V{\"o}gler}, {Katsukawa}, \&
  {Tsuneta}}]{2008A&A...484L..17D}
{Danilovic}, S., {Gandorfer}, A., {Lagg}, A., {et~al.} 2008, \aap, 484, L17

\bibitem[{{Danilovic} {et~al.}(2016{\natexlab{a}}){Danilovic}, {Rempel}, {van
  Noort}, \& {Cameron}}]{2016A&A...594A.103D}
{Danilovic}, S., {Rempel}, M., {van Noort}, M., \& {Cameron}, R.
  2016{\natexlab{a}}, \aap, 594, A103

\bibitem[{{Danilovic} {et~al.}(2016{\natexlab{b}}){Danilovic}, {van Noort}, \&
  {Rempel}}]{2016A&A...593A..93D}
{Danilovic}, S., {van Noort}, M., \& {Rempel}, M. 2016{\natexlab{b}}, \aap,
  593, A93

\bibitem[{{de la Cruz Rodr{\'{\i}}guez} {et~al.}(2016){de la Cruz
  Rodr{\'{\i}}guez}, {Leenaarts}, \& {Asensio Ramos}}]{2016ApJ...830L..30D}
{de la Cruz Rodr{\'{\i}}guez}, J., {Leenaarts}, J., \& {Asensio Ramos}, A.
  2016, \apjl, 830, L30

\bibitem[{{de la Cruz Rodr{\'{\i}}guez} {et~al.}(2019){de la Cruz
  Rodr{\'{\i}}guez}, {Leenaarts}, {Danilovic}, \&
  {Uitenbroek}}]{2019A&A...623A..74D}
{de la Cruz Rodr{\'{\i}}guez}, J., {Leenaarts}, J., {Danilovic}, S., \&
  {Uitenbroek}, H. 2019, \aap, 623, A74

\bibitem[{{De Pontieu} {et~al.}(2014){De Pontieu}, {Title}, {Lemen}, {Kushner},
  {Akin}, {Allard}, {Berger}, {Boerner}, {Cheung}, {Chou}, {Drake}, {Duncan},
  {Freeland}, {Heyman}, {Hoffman}, {Hurlburt}, {Lindgren}, {Mathur}, {Rehse},
  {Sabolish}, {Seguin}, {Schrijver}, {Tarbell}, {W{\"u}lser}, {Wolfson},
  {Yanari}, {Mudge}, {Nguyen-Phuc}, {Timmons}, {van Bezooijen}, {Weingrod},
  {Brookner}, {Butcher}, {Dougherty}, {Eder}, {Knagenhjelm}, {Larsen},
  {Mansir}, {Phan}, {Boyle}, {Cheimets}, {DeLuca}, {Golub}, {Gates}, {Hertz},
  {McKillop}, {Park}, {Perry}, {Podgorski}, {Reeves}, {Saar}, {Testa}, {Tian},
  {Weber}, {Dunn}, {Eccles}, {Jaeggli}, {Kankelborg}, {Mashburn}, {Pust},
  {Springer}, {Carvalho}, {Kleint}, {Marmie}, {Mazmanian}, {Pereira}, {Sawyer},
  {Strong}, {Worden}, {Carlsson}, {Hansteen}, {Leenaarts}, {Wiesmann},
  {Aloise}, {Chu}, {Bush}, {Scherrer}, {Brekke}, {Martinez-Sykora}, {Lites},
  {McIntosh}, {Uitenbroek}, {Okamoto}, {Gummin}, {Auker}, {Jerram}, {Pool}, \&
  {Waltham}}]{2014SoPh..289.2733D}
{De Pontieu}, B., {Title}, A.~M., {Lemen}, J.~R., {et~al.} 2014, \solphys, 289,
  2733

\bibitem[{{Esteban Pozuelo} {et~al.}(2019){Esteban Pozuelo}, {de la Cruz
  Rodr{\'\i}guez}, {Drews}, {Rouppe van der Voort}, {Scharmer}, \&
  {Carlsson}}]{2019ApJ...870...88E}
{Esteban Pozuelo}, S., {de la Cruz Rodr{\'\i}guez}, J., {Drews}, A., {et~al.}
  2019, \apj, 870, 88

\bibitem[{Guennebaud {et~al.}(2010)Guennebaud, Jacob, {et~al.}}]{eigenweb}
Guennebaud, G., Jacob, B., {et~al.} 2010, Eigen v3, http://eigen.tuxfamily.org

\bibitem[{{Joshi} {et~al.}(2011){Joshi}, {Pietarila}, {Hirzberger}, {Solanki},
  {Aznar Cuadrado}, \& {Merenda}}]{2011ApJ...734L..18J}
{Joshi}, J., {Pietarila}, A., {Hirzberger}, J., {et~al.} 2011, \apjl, 734, L18

\bibitem[{{Lagg} {et~al.}(2014){Lagg}, {Solanki}, {van Noort}, \&
  {Danilovic}}]{2014A&A...568A..60L}
{Lagg}, A., {Solanki}, S.~K., {van Noort}, M., \& {Danilovic}, S. 2014, \aap,
  568, A60

\bibitem[{{Leenaarts} {et~al.}(2018){Leenaarts}, {de la Cruz Rodr{\'\i}guez},
  {Danilovic}, {Scharmer}, \& {Carlsson}}]{2018A&A...612A..28L}
{Leenaarts}, J., {de la Cruz Rodr{\'\i}guez}, J., {Danilovic}, S., {Scharmer},
  G., \& {Carlsson}, M. 2018, \aap, 612, A28

\bibitem[{{Lites} {et~al.}(1990){Lites}, {Scharmer}, \&
  {Skumanich}}]{1990ApJ...355..329L}
{Lites}, B.~W., {Scharmer}, G.~B., \& {Skumanich}, A. 1990, \apj, 355, 329

\bibitem[{{L{\"o}fdahl}(2002)}]{2002SPIE.4792..146L}
{L{\"o}fdahl}, M.~G. 2002, in Society of Photo-Optical Instrumentation
  Engineers (SPIE) Conference Series, Vol. 4792, Image Reconstruction from
  Incomplete Data, ed. P.~J. {Bones}, M.~A. {Fiddy}, \& R.~P. {Millane},
  146--155

\bibitem[{{Oba} {et~al.}(2017){Oba}, {Riethm{\"u}ller}, {Solanki}, {Iida},
  {Quintero Noda}, \& {Shimizu}}]{2017ApJ...849....7O}
{Oba}, T., {Riethm{\"u}ller}, T.~L., {Solanki}, S.~K., {et~al.} 2017, \apj,
  849, 7

\bibitem[{{Orozco Su{\'a}rez} \& {Del Toro
  Iniesta}(2007)}]{2007A&A...462.1137O}
{Orozco Su{\'a}rez}, D. \& {Del Toro Iniesta}, J.~C. 2007, \aap, 462, 1137

\bibitem[{{Piskunov} \& {Kochukhov}(2002)}]{2002A&A...381..736P}
{Piskunov}, N. \& {Kochukhov}, O. 2002, \aap, 381, 736

\bibitem[{{Piskunov} {et~al.}(1995){Piskunov}, {Kupka}, {Ryabchikova}, {Weiss},
  \& {Jeffery}}]{1995A&AS..112..525P}
{Piskunov}, N.~E., {Kupka}, F., {Ryabchikova}, T.~A., {Weiss}, W.~W., \&
  {Jeffery}, C.~S. 1995, \aaps, 112, 525

\bibitem[{{Piskunov} {et~al.}(1990){Piskunov}, {Tuominen}, \&
  {Vilhu}}]{1990A&A...230..363P}
{Piskunov}, N.~E., {Tuominen}, I., \& {Vilhu}, O. 1990, \aap, 230, 363

\bibitem[{{Quintero Noda} {et~al.}(2016{\natexlab{a}}){Quintero Noda},
  {Shimizu}, {Ruiz Cobo}, {Suematsu}, {Katsukawa}, \&
  {Ichimoto}}]{2016MNRAS.460.1476Q}
{Quintero Noda}, C., {Shimizu}, T., {Ruiz Cobo}, B., {et~al.}
  2016{\natexlab{a}}, \mnras, 460, 1476

\bibitem[{{Quintero Noda} {et~al.}(2016{\natexlab{b}}){Quintero Noda},
  {Suematsu}, {Ruiz Cobo}, {Shimizu}, \& {Asensio Ramos}}]{2016MNRAS.460..956Q}
{Quintero Noda}, C., {Suematsu}, Y., {Ruiz Cobo}, B., {Shimizu}, T., \&
  {Asensio Ramos}, A. 2016{\natexlab{b}}, \mnras, 460, 956

\bibitem[{{Rempel}(2012)}]{2012ApJ...750...62R}
{Rempel}, M. 2012, \apj, 750, 62

\bibitem[{{Riethm{\"u}ller} \& {Solanki}(2019)}]{2019A&A...622A..36R}
{Riethm{\"u}ller}, T.~L. \& {Solanki}, S.~K. 2019, \aap, 622, A36

\bibitem[{{Ruiz Cobo} \& {Asensio Ramos}(2013)}]{2013A&A...549L...4R}
{Ruiz Cobo}, B. \& {Asensio Ramos}, A. 2013, \aap, 549, L4

\bibitem[{{Ryabchikova} {et~al.}(2015){Ryabchikova}, {Piskunov}, {Kurucz},
  {Stempels}, {Heiter}, {Pakhomov}, \& {Barklem}}]{2015PhyS...90e4005R}
{Ryabchikova}, T., {Piskunov}, N., {Kurucz}, R.~L., {et~al.} 2015, \physscr,
  90, 054005

\bibitem[{{Scharmer} {et~al.}(2003){Scharmer}, {Bjelksjo}, {Korhonen},
  {Lindberg}, \& {Petterson}}]{2003SPIE.4853..341S}
{Scharmer}, G.~B., {Bjelksjo}, K., {Korhonen}, T.~K., {Lindberg}, B., \&
  {Petterson}, B. 2003, in \procspie, Vol. 4853, Innovative Telescopes and
  Instrumentation for Solar Astrophysics, ed. S.~L. {Keil} \& S.~V. {Avakyan},
  341--350

\bibitem[{{Scharmer} {et~al.}(2013){Scharmer}, {de la Cruz Rodriguez},
  {S{\"u}tterlin}, \& {Henriques}}]{2013A&A...553A..63S}
{Scharmer}, G.~B., {de la Cruz Rodriguez}, J., {S{\"u}tterlin}, P., \&
  {Henriques}, V.~M.~J. 2013, \aap, 553, A63

\bibitem[{{Scharmer} {et~al.}(2011){Scharmer}, {Henriques}, {Kiselman}, \& {de
  la Cruz Rodr{\'\i}guez}}]{2011Sci...333..316S}
{Scharmer}, G.~B., {Henriques}, V.~M.~J., {Kiselman}, D., \& {de la Cruz
  Rodr{\'\i}guez}, J. 2011, Science, 333, 316

\bibitem[{{Scharmer} {et~al.}(2019){Scharmer}, {L{\"o}fdahl}, {Sliepen}, \& {de
  la Cruz Rodr{\'\i}guez}}]{2019A&A...626A..55S}
{Scharmer}, G.~B., {L{\"o}fdahl}, M.~G., {Sliepen}, G., \& {de la Cruz
  Rodr{\'\i}guez}, J. 2019, Astronomy and Astrophysics, 626, A55

\bibitem[{{Scharmer} {et~al.}(2010){Scharmer}, {L{\"o}fdahl}, {van Werkhoven},
  \& {de la Cruz Rodr{\'\i}guez}}]{2010A&A...521A..68S}
{Scharmer}, G.~B., {L{\"o}fdahl}, M.~G., {van Werkhoven}, T.~I.~M., \& {de la
  Cruz Rodr{\'\i}guez}, J. 2010, Astronomy and Astrophysics, 521, A68

\bibitem[{Tikhonov \& Arsenin(1977)}]{Tikhonov77}
Tikhonov, A.~N. \& Arsenin, V.~Y. 1977, Solutions of Ill-posed problems
  (W.H.~Winston)

\bibitem[{{Tiwari} {et~al.}(2013){Tiwari}, {van Noort}, {Lagg}, \&
  {Solanki}}]{2013A&A...557A..25T}
{Tiwari}, S.~K., {van Noort}, M., {Lagg}, A., \& {Solanki}, S.~K. 2013, \aap,
  557, A25

\bibitem[{{van Noort}(2012)}]{2012A&A...548A...5V}
{van Noort}, M. 2012, \aap, 548, A5

\bibitem[{{van Noort} {et~al.}(2013){van Noort}, {Lagg}, {Tiwari}, \&
  {Solanki}}]{2013A&A...557A..24V}
{van Noort}, M., {Lagg}, A., {Tiwari}, S.~K., \& {Solanki}, S.~K. 2013, \aap,
  557, A24

\bibitem[{{van Noort} {et~al.}(2005){van Noort}, {Rouppe van der Voort}, \&
  {L{\"o}fdahl}}]{2005SoPh..228..191V}
{van Noort}, M., {Rouppe van der Voort}, L., \& {L{\"o}fdahl}, M.~G. 2005,
  Solar Physics, 228, 191

\bibitem[{{Vissers} {et~al.}(2019){Vissers}, {de la Cruz Rodr{\'\i}guez},
  {Libbrecht}, {Rouppe van der Voort}, {Scharmer}, \&
  {Carlsson}}]{2019A&A...627A.101V}
{Vissers}, G.~J.~M., {de la Cruz Rodr{\'\i}guez}, J., {Libbrecht}, T., {et~al.}
  2019, \aap, 627, A101

\bibitem[{{V{\"o}gler} {et~al.}(2005){V{\"o}gler}, {Shelyag}, {Sch{\"u}ssler},
  {Cattaneo}, {Emonet}, \& {Linde}}]{2005A&A...429..335V}
{V{\"o}gler}, A., {Shelyag}, S., {Sch{\"u}ssler}, M., {et~al.} 2005, \aap, 429,
  335

\end{thebibliography}

\end{document}